\documentclass[
    reprint,
    preprintnumbers,
    superscriptaddress,
    nofootinbib,
     amsmath,amssymb,
     aps,
     prd,
    floatfix,
    longbibliography]{revtex4-2}
    
    \usepackage{graphicx}
    \usepackage[caption=false]{subfig}
    \usepackage[usenames,dvipsnames]{xcolor} 
    \usepackage{pgfplots}
    \usepackage[utf8]{inputenc}
    \usepackage{color}
    \usepackage{hyperref}
    \usepackage[normalem]{ulem} 
    \usepackage{physics}
    \usepackage{enumitem}
    \usepackage{comment}
    \usepackage{bm}
    \usepackage{aas_macros}
    \usepackage[capitalise]{cleveref}
\hypersetup{
  breaklinks = true,
  colorlinks   = true, 
  urlcolor     = blue, 
  linkcolor    = blue, 
  citecolor   = blue 
}
    \allowdisplaybreaks[1]
    \graphicspath{{figures/}}

\usepackage{float}

\newcommand{\Cov}{{\rm Cov}}
\newcommand{\md}{{\rm d}}

\newcommand{\oxford}{Astrophysics, University of Oxford, DWB, Keble Road, Oxford OX1 3RH, United Kingdom}

\newcommand{\iap}{CNRS \& Sorbonne Universit\'{e}, Institut d’Astrophysique de Paris (IAP), UMR 7095, 98 bis bd Arago, F-75014 Paris, France}

\newcommand{\splitatcommas}[1]{%
  \begingroup
  \begingroup\lccode`~=`, \lowercase{\endgroup
    \edef~{\mathchar\the\mathcode`, \penalty0 \noexpand\hspace{0pt plus 1em}}%
  }\mathcode`,="8000 #1%
  \endgroup
}

\begin{document}

\title{Scant evidence for thawing quintessence}

\author{William J. Wolf}
\email{william.wolf@stx.ox.ac.uk}
\affiliation{\oxford}
\author{Carlos Garc\'ia-Garc\'ia}
\email{carlos.garcia-garcia@physics.ox.ac.uk}
\affiliation{\oxford}
\author{Deaglan J. Bartlett}
\email{deaglan.bartlett@iap.fr}
\affiliation{\iap}
\author{Pedro G. Ferreira}
\email{pedro.ferreira@physics.ox.ac.uk}
\affiliation{\oxford}

\begin{abstract}
New constraints on the expansion rate of the Universe seem to favor evolving dark energy in the form  of thawing quintessence models, i.e., models for which a canonical, minimally coupled scalar field has, at late times, begun to evolve away from potential energy domination. We scrutinize the evidence for thawing quintessence by exploring what it predicts for the equation of state. We show that, in terms of the usual Chevalier-Polarski-Linder parameters, ($w_0$, $w_a$), thawing quintessence is, in fact, only marginally consistent with a compilation of the current data. Despite this, we embrace the possibility that thawing quintessence is dark energy and find constraints on the microphysics of 
this scenario. We do so 
in terms of the effective mass $m^2$ and energy scale $V_0$ of the scalar field potential. We are particularly careful to enforce un-informative, flat priors on these parameters so as to minimize their effect on the final posteriors. While the current data favors a large and negative value of $m^2$, when we compare these models to the standard $\Lambda$CDM model we find that there is scant evidence for thawing quintessence. 
\end{abstract}

\maketitle

\section{Introduction}
A combination of the latest measurements of baryon acoustic oscillations (BAO) \cite{DESI:2024mwx}, luminosity distances of distant supernovae (SNe) \cite{Rubin:2023ovl, Scolnic:2021amr, DES:2024tys}, and  measurements of the Cosmic Microwave Background (CMB) \cite{Planck:2018vyg, Planck:2019nip, ACT:2023dou, ACT:2023kun} have provided  hints of  deviations from the $\Lambda$-Cold Dark Matter ($\Lambda$CDM) paradigm. One possible interpretation is that the accelerated expansion of the Universe at late times is due to some form of evolving dark energy, such as a quintessence scalar field, with a time evolving equation of state.

Much of the discussion in the literature has focused on the Chevallier-Polarski-Linder (CPL) parameterization \cite{Linder:2002et, Chevallier:2000qy} of the equation of state of dark energy, $w\equiv P_{\varphi}/\rho_{\varphi}$ (where $P_{\varphi}$ and $\rho_{\varphi}$ are the pressure and energy density of the dark energy), in which
\begin{equation}\label{param}
w(a)=w_0+w_a(1-a),
\end{equation}
where $a$ is the scale factor of the Universe.
$w_0$ gives the value of $w$ today and $w_a$ characterizes its temporal evolution. 
$\Lambda$CDM is given by $w_a = 0$ and $w_0 = -1$, while a variety of dynamically driven dark energy possibilities occupy the rest of the parameter space defined by the ($w_0$, $w_a$) plane. 

\begin{figure}
    \centering
    \includegraphics[width=\columnwidth]{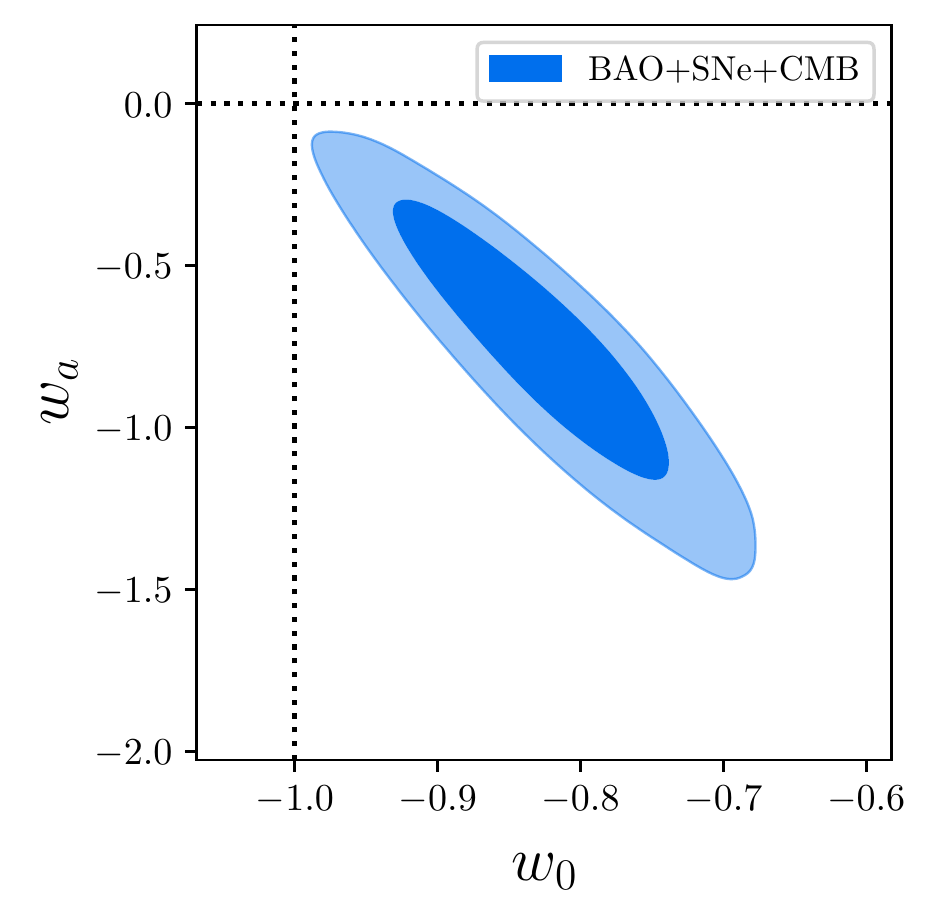}
    \caption{Constraints on CPL parameters (68\% and 95\% C.L.) using the combined DESI BAO data \cite{DESI:2024mwx}, Pantheon+ SNe data \cite{Scolnic:2021amr}, and CMB (Planck and ACT lensing) data \cite{Planck:2018vyg, Planck:2019nip,ACT:2023kun, ACT:2023dou}.
    }
    \label{fig: DESI contour}
\end{figure}

The CPL parametrization is somewhat of a blunt instrument, a form of data compression which is amenable to over-interpretation \cite{Wolf:2023uno}. Nevertheless, the constraints on $w_0$ and $w_a$ seem to tell an intriguing story about the microphysics of dark energy as is apparent on close inspection of \cref{fig: DESI contour}. For a start, they seem to show that $w_a\neq 0$, which would represent evidence for a time evolving equation of state. Furthermore, the locale in the ($w_0$, $w_a$) plane which is singled out by the current data seems to favour ``thawing'' models of dark energy, i.e., models which will have started off with $w(a)\simeq-1$ before evolving to the larger values that they have now. And finally, the inclination of the degeneracy of the ($w_0$, $w_a$) constraints penetrates any area corresponding to ``phantom'' dark energy, i.e., dark energy in which the equation of state is $w<-1$. 

While the new constraints on ($w_0$, $w_a$) are not yet definitive, with particular choices of data sets being questioned in the analysis \cite{Efstathiou:2024dvn,Efstathiou_2024}, their qualitative trend has led to an avalanche of results concerning dark energy \cite{Tada:2024znt, Berghaus:2024kra, Park:2024jns, Yin:2024hba, Shlivko:2024llw, Cortes:2024lgw, DESI:2024kob, Dinda:2024kjf, Carloni:2024zpl, Wang:2024rjd, Croker:2024jfg, Mukherjee:2024ryz, Roy:2024kni, Wang:2024dka, Wang:2024hwd, Gialamas:2024lyw, Notari:2024rti, Orchard:2024bve, Wang:2024sgo, Giare:2024gpk, Ye:2024ywg, Dinda:2024ktd, Jiang:2024xnu, Ghosh:2024kyd, Luongo:2024fww, Alfano:2024jqn, Reboucas:2024smm, Pang:2024qyh, Ramadan:2024kmn, Heckman:2024apk, Bhattacharya:2024hep, DESI:2024aqx, Wolf:2024stt, Arjona:2024dsr}. We highlight a few important  comments and results that have arisen. First of all, and following the point made in \citet{Wolf:2023uno}, a few authors argued that the ($w_0$, $w_a$) determined from a given survey are tied to the specific properties of {\it that} survey. In other words, while ($w_0$, $w_a$) are, naively, often interpreted as the first terms of a Taylor expansion of the equation of state, they are in fact fitting parameters which depend on the survey properties. \citet{Cortes:2024lgw} made the further point that the CPL parameterization was in fact showing that $w\simeq-1$ in the redshift range where there is most data. \citet{Shlivko:2024llw} echoed the point made in \citet{Wolf:2023uno}, that one cannot infer the microphysics of the dark energy model from the inferred ($w_0$, $w_a$) parameters and, for example, whether it is phantom or not because this involves extrapolating the best fit $w(a)$ far outside the redshifts at which the data are used to determine the parameters. On the other hand, more general, model agnostic studies such as \citet{Ye:2024ywg}, \citet{DESI:2024kob}, and \citet{DESI:2024aqx} point towards a (marginally) statistically significant detection of phantom-like behavior for $z>0.6$, where $z$ is the cosmological redshift.

In \cite{Wolf:2023uno}, it was argued that a particularly simple quintessence model, with a scalar field, $\varphi$, and a potential 
\begin{equation}\label{potential}
    V (\varphi) = V_0 + \frac{1}{2}m^2\varphi^2,
\end{equation}
can, in principle, cover an incredibly wide region of the ``thawing'' quadrant  of the $(w_0, w_a)$ plane. In essence, this potential represents the leading order terms in an effective field theory expansion for a huge variety of scalar field models in the region of field space in which it can have an impact on the dynamics of the Universe. Thus, it was argued, the vast majority of dark energy models under consideration (i.e., thawing models) are undetermined. In other words, with the types of cosmological measurement available, it will be impossible to identify the specific microphysics of the dark energy model, beyond that given by \cref{potential}.

In this paper, we study what the new constraints on the expansion rate of the Universe tell us about quintessence. More specifically, we directly map the dark energy model given by \cref{potential} into the ($w_0$, $w_a$) plane considering the data provided by BAO, SNe, and CMB measurements. We find that, from the point of view of the CPL parameterization, and once all of the data is accounted for, thawing quintessence (as represented by \cref{potential}) is not favored by the BAO, SNe and CMB data. Thus, thawing quintessence more generally lies mostly outside the favoured regions once it is directly fit to the data. Nevertheless, we can still examine the implications for thawing quintessence, and to this end, we directly constrain the microphysics of thawing quintessence in terms of $V_0$ and $m^2$.

The paper proceeds as follows. In \cref{Sec: Dynamical Dark Energy} we review dynamical dark energy and thawing quintessence, and describe the different regimes it has, depending on whether $m^2$ is negative or positive. This allows us to understand how the potential from \cref{potential} is able to cover such a broad range of thawing behaviour.
In \cref{sec:CPLthawing} we describe, in detail, how thawing quintessence maps onto the ($w_0$, $w_a$) plane, and how this mapping is dependent on the survey characteristics one is using to measure the CPL parameters. 
We show how, in the case that we consider a compilation of all data sets (including, in particular, the CMB) thawing quintessence is only marginally consistent with the current constraints on ($w_0$, $w_a$).
In \cref{Sec:MCMCresults}, we find constraints on the microphysics of thawing quintessence, taking particular care to understand the priors on $V_0$ and $m^2$, and how they affect the posteriors we obtain. An important, novel aspect, of the analysis is our use of an analytic ansatz to find the initial conditions for the scalar field evolution, using Symbolic Regression techniques described in \cref{sec.symbolic}. 
In \cref{sec:conclusion} we discuss the implications of our findings.
In a series of Appendices we lay out some of the details and, more significantly, new techniques developed for this analysis.

\section{Thawing Dark Energy}\label{Sec: Dynamical Dark Energy}

In this paper we will focus on dark energy driven by a quintessence scalar field and described by the action:
\begin{equation}
S=\int d^4 x \sqrt{-g}\left[\frac{1}{2} M_{\mathrm{\rm Pl}}^2 R-\frac{1}{2} g^{\mu \nu} \partial_\mu \varphi \partial_\nu \varphi-V(\varphi)\right]+S_{\rm m}.
\label{sfaction}
\end{equation}
$M_{\rm Pl}$ is the reduced Planck mass, $g$ is the determinant of the metric $g_{\mu\nu}$, $R$ is the Ricci curvature scalar, $V(\varphi)$ is the scalar field potential, and $S_{\rm m}$ is the action for matter. The scalar field has a canonical kinetic term and is minimally coupled to gravity through the metric determinant.

Such a theory leads to a  dark energy equation of state $w(a)$,
\begin{equation}\label{DEeq}
 w(a) =\frac{\frac{\dot{\varphi}^2}{2}-V(\varphi)}{\frac{\dot{\varphi}^2}{2}+V(\varphi)},
\end{equation}
which evolves in time with the evolution of the scalar field,
where overdots are derivatives with respect to cosmic time.
The evolution of the scalar field is given by the scalar field equation of motion,
\begin{equation}\label{ScalarEOM}
    \ddot{\varphi} + 3 H \dot{\varphi} + V'(\varphi) = 0,
\end{equation}
where $V'(\varphi) \equiv dV/d\varphi$ and $H$ is the expansion rate of the Universe given by the first Friedmann equation,
\begin{equation}\label{Friedmann}
H^2\equiv 
\left(\frac{\dot{a}}{a}\right)^2=\frac{1}{3M^2_{pl}} (\rho+\rho_{\varphi}),
\end{equation}
where $\rho_\varphi$ and $\rho$ are, respectively, the scalar field and ordinary matter energy densities and the Hubble parameter today is $H_0=100 h$ km s$^{-1}$ Mpc$^{-1}$ with $h\simeq 0.67$.

The behavior of  $w(a)$ is the most straightforward way to characterize dark energy, and one can identify different broad classes of dark energy models, such as thawing (when $dw/da>0$) or freezing (when $dw/da<0$) models \cite{Caldwell:2005tm}.  In the case of a quintessence scalar field with a canonical kinetic energy, as in \cref{sfaction}, the dynamical behavior of $w(a)$ is largely determined by the potential $V(\varphi)$. While a great number of potentials have been considered in the literature, the physics of thawing quintessence can be effectively captured by \cref{potential} where, throughout this paper, $m^2$ will be in units of $(H_0/h)^{2}$ and $V_0$ in units of $M^2_{\rm Pl}(H_0/h)^{2}$.
This has do with the fact that any arbitrary analytic potential can be represented by an effective Taylor expansion in $\varphi$,
combined with a field redefinition to get rid of the linear term \cite{Wolf:2023uno}.
For example, exponential tracker potentials \cite{Ferreira:1997hj, Copeland:1997et, Caldwell:1997ii} are well-described at leading order by the standard ($m^2>0$) quadratic version of \cref{potential}, while pNGB and axion potentials \cite{Frieman:1995pm, Marsh:2015xka, Dutta:2008qn} are well-described by the hilltop ($m^2<0$) version of \cref{potential} (see e.g., \cite{Dutta:2008qn, Wolf:2023uno, Boubekeur:2005zm, Wolf:2024lbf} for more on the quadratic hilltop potential). Consequently, the potential given in \cref{potential} can be understood as the leading order description for many potentials of great theoretical interest, and characterizes all of these distinct theories in terms of an energy scale $V_0$ and an effective mass $m^2 = d^2V/d\varphi^2$ in the leading order quadratic term.

Furthermore, the potential of \cref{potential} captures the entire phenomenology of thawing quintessence if one thinks of it, purely, in terms of $w(a)$. The $m^2 > 0$ branch of the model phenomenologically captures standard ``slow-roll'' thawing quintessence, where all of these models converge to a highly linear, universal dynamical behavior in the dark energy equation of state $w(a)$ given by $dw/da|_{a=1} \approx 1.5 [1+ w(a=1)]$ (see e.g., \cite{Garcia-Garcia:2019cvr, Scherrer:2007pu, Linder:2015zxa, Scherrer:2015tra, Marsh:2014xoa, Wolf:2023uno}). On the other hand, the $m^2 < 0$ ``hilltop'' branch behaves quite differently as there are an exceedingly wide range of dynamically possible behaviors for $w(a)$. For example, at small $\left|m^2\right|$, the hilltop models approach the linear evolution of the $m^2 > 0$ models, while at larger $\left|m^2\right|$ the hilltop models can evolve in a highly non-linear manner, where $w(a)$ is flat across most of the Universe's history but shoots sharply upwards at more recent redshifts (see \cite{Dutta:2008qn, Chiba:2009sj, Wolf:2023uno, Shlivko:2024llw} for further discussion). Consequently, this model can in principle be mapped across the $(w_0, w_a)$ plane depending on the choices of mass, energy scale, and initial conditions \cite{Wolf:2023uno, Shlivko:2024llw}. Thus, the potential of \cref{potential} is an ideal model to consider constraints on thawing quintessence microphysics in terms of an energy scale and effective mass. 

It is useful to understand the dynamics of thawing quintessence driven by the potential in \cref{potential} in slightly more detail. Without loss of generality, we will assume that ${\dot \varphi}= 0$ initially. If $m^2>0$, we have that $\varphi$ will start off with $\varphi_{\rm ini}\neq 0$ and a potential energy given by $V=V_0+\frac{1}{2}m^2\varphi_{\rm ini}^2$. Initially, the scalar field will remain frozen due to Hubble friction, slowly developing a kinetic energy over time. Once the scalar field has sped up enough, $w(a)$ will start to deviate from $-1$, and, at some point, the scalar field will end up oscillating around the minimum of the potential. For time scales such that $mt\gg 1$, the equation of state is that of dust, $w(a)\simeq 0$ \cite{Turner:1983he}, but for the choice of parameters we will consider here, that is well beyond the time scales we will be looking at. Note that, for $m^2>0$, we will still recover a positive energy density and $w(a)=-1$ even if we set $V_0=0$. With both $V_0$ and $m^2>0$, we expect to have a degeneracy at early times between the two parameters. Indeed, we can define  the effective cosmological constant at early times to be $\Lambda=(V_0+\frac{1}2{m^2}\varphi_{\rm ini})/(3M^2_{\rm Pl})$, making manifest the degeneracy between the two parameters. This degeneracy will play an important role in the understanding the posterior constraints on thawing quintessence in \cref{Sec:MCMCresults}.

If $m^2<0$, the situation is markedly different \cite{Dutta:2008qn, Chiba:2009sj, Wolf:2023uno}. 
The potential is now a ``hilltop'' and the solutions are unstable, exponential in $|m|t$. The larger $|m|$, the faster the instability. Now, one can now tune the initial conditions, placing $\varphi_{\rm ini}$ arbitrarily close to $0$ leading to an extended period in which $w(a)=-1$. But once the scalar field has grown enough, the onset of kinetic energy domination can be rapid. This translates in an equation of state which remains at $-1$ over most of the Universe's expansion history and then rapidly transitions to $w>-1$. The higher the mass, the later, but also more rapid, the transition away from $w\sim-1$.

The case of $m^2=0$ merits a brief note. We are considering the case where ${\dot \varphi}_{\rm ini}=0$. If that is the case, and $m^2=0$, we have that $V=V_0$ throughout its evolution and we recover the case of a cosmological constant. We can assume that ${\dot \varphi}_{\rm ini}\neq0$ but we then have $\dot\varphi\propto a^{-3}$ and is thus very rapidly redshifting away. So we have that $m^2=0$ effectively corresponds to $w(a)=-1$ throughout.

\section{Thawing quintessence from the point of view of ($w_0$, $w_a$).}\label{sec:CPLthawing}

\begin{figure}
    \includegraphics[width=\columnwidth]{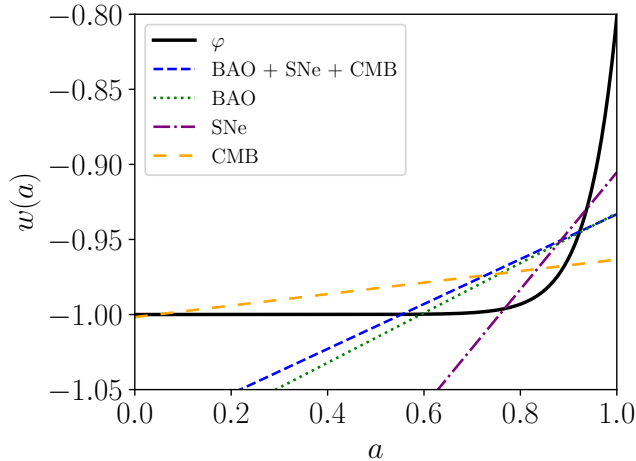}
    \caption{Equation of state, $w(a)$, for a scalar field model with $m^2<0$ ($(V_0, m^2) = (0.9725, -40.0)$, black solid line) alongside the corresponding best fit CPL models when using the BAO, SNe, and CMB survey characteristics (as described in \cref{sec:CPLthawing}) both individually and combined, showing that the best fit ($w_0$, $w_a$) parameters are clearly sensitive to the survey parameters.
    }\label{Fig:fit_model}
\end{figure}

In this section we explore how thawing quintessence maps on to the ($w_0$, $w_a$) plane of the CPL parametrization.
To do so, we need to recap how ($w_0$, $w_a$) are actually inferred from cosmological data.
In practice, we do not directly measure $w(a)$, but infer ($w_0$, $w_a$) from data that probe $H(z)$ at different redshifts.  In the CPL parameterization, $H$ is given by 
\begin{equation}\label{hfit}
    H^2(a)=H_0^2\left[\Omega_{\mathrm{m}} a^{-3}+\left(1-\Omega_{\mathrm{m}}\right) e^{3 w_a (a-1)} a^{-3(1+w_0+w_a)}\right],
\end{equation}
where $\Omega_{\mathrm{m}}$ is the fractional energy density in non-relativistic matter today and $H_0$ is the Hubble constant.
A dark energy model's representation in terms of ($w_0$, $w_a$)  depends on how  \cref{hfit} compares with actual measurements of cosmological quantities sensitive to $H(z)$.

We now map all of thawing quintessence as given by \cref{potential} into the ($w_0$, $w_a$) plane given by the CPL parameterization. We do so by mimicking what is done with real data. For a given choice of ($V_0$, $m^2$) we generate the associated $H(z)$ and then the equivalent observables which are measured when one uses BAO, SNe and CMB data. Given these observables we then find the best-fit CPL parameters. 
In this paper we will focus on the BAO data from DESI \cite{DESI:2024mwx}, the SNe data from Pantheon+ \cite{Scolnic:2021amr}, and the CMB data from Planck and ACT \cite{Planck:2018vyg, Planck:2019nip, ACT:2023dou, ACT:2023kun}.

To be more specific we determine the corresponding CPL parameters, ($w_0$, $w_a$), given the characteristics of a particular survey, ``S''. We characterize a given survey in terms of a set of observables ${\mathcal O}_i$ at different redshifts, $z_i$, and their associated covariance, $\Cov$.
We construct the distance between two models in terms of 
\begin{equation}\label{chi2}
\chi^2_{\rm S} = \left(\mathcal{O}^{\rm CPL} - \mathcal{O}^{\varphi} \right)^T \left(\frac{\mathcal{O}^{\rm data}}{\mathcal{O}^{\varphi}}\right)^T {\Cov}^{-1} \frac{\mathcal{O}^{\rm data}}{\mathcal{O}^{\varphi}} \left(\mathcal{O}^{\rm CPL} - \mathcal{O}^{\varphi} \right),
\end{equation}
where $\mathcal{O}^{\rm CPL}$ and $\mathcal{O}^{\varphi}$ are the observables computed with the CPL parameterization ($w_0$, $w_a$), or the scalar field ($V_0$, $m^2$), respectively, and the factors $\mathcal{O}^{\rm data} / \mathcal{O}^{\varphi}$ ensure that we fit the observables using the surveys measurements' relative errors, and not the absolute ones.  Thus, for a set of thawing quintessence parameters, ($V_0$, $m^2$), we minimize \cref{chi2} to find the corresponding CPL parameters, ($w_0$, $w_a$). So, for example, mapping ($V_0$, $m^2$) into the CPL parameters ($w_0$, $w_a$) for the DESI survey would involve minimizing $\chi^2_{\rm DESI}$ from \cref{chi2}, while doing so for a combination of data sets  can be done by minimizing $\chi^2_{\rm tot} = \chi^2_{\rm DESI} + \chi^2_{\rm SNe} + \chi^2_{\rm CMB}$.

For this process to be at all feasible, we consider a reduced or compressed version of the data.
In \cref{data}, we explain how the data can be compressed into a set of ``measured'' physical parameters and their covariance. In brief, the CMB data can be compressed into an acoustic scale $\ell_{\mathrm{A}}$ and shift parameter $R\left(z_{*}\right)$ at the photon decoupling redshift $z_{*}$ \cite{Komatsu_2009,Planck:2015bue,Chen:2018dbv},
the SNe data are sensitive to the luminosity distance {and can be compressed into measurements of $E(z) \equiv H(z) / H_0$ at different redshifts \cite{Riess:2017lxs}}, and the BAO data are {directly} sensitive to the angular diameter distance or Hubble parameter, which can be directly used in our fit (see \cref{data} for details).

In \cite{Wolf:2023uno} we showed that, due to wide range of dynamical behavior in the evolution of $w(a)$ in some of these models, the exact representation of these models in terms of ($w_0$, $w_a$) can potentially depend sensitively on which redshifts are probed in distance measurements sensitive to $H(z)$. In other words, due to the potential non-linearity in the evolution of the equation of state for this model, the best fit linear CPL parameterization can look different depending on which redshifts the surveys are probing and thus which data points and observations are considered in the fitting procedure. 

To illustrate this point, consider the  example in \cref{Fig:fit_model} where we depict the cosmological evolution of the equation of state for a particular (and somewhat generic) thawing model from the $m^2 < 0$ branch of \cref{potential} with parameters $(V_0, m^2) = (0.9725, -40.0)$, as well as its corresponding best fit CPL models for various data sets, echoing what we have said in \cite{Wolf:2023uno}. In this article, we are considering data from BAO, SNe, and CMB observations, all of which can be understood to be sensitive to $H(z)$ at different redshift epochs. The best fit ($w_0$, $w_a$) can differ quite substantially depending on the redshifts the observables are sensitive to, with the Pantheon+ SNe \cite{Scolnic:2021amr} data probing redshifts up to $z \simeq 0.07$ giving ($w_0$, $w_a$) $\simeq$ ($-0.91$, $-0.40$), DESI BAO data probing redshifts up to $z \simeq 0.295$ \cite{DESI:2024mwx} giving ($w_0$, $w_a$) $\simeq$ ($-0.93$, $-0.17$), CMB data locked in at the photon decoupling epoch $ z \simeq 1090$ \cite{Planck:2018vyg} giving ($w_0$, $w_a$) $\simeq$ ($-0.96$, $-0.04$), and the combined use of all the data sets giving ($w_0$, $w_a$)$\simeq$ ($-0.93$, $-0.15$). 

The more recent the epoch probed, the steeper the region inferred in the $(w_0, w_a)$ plane, which is due to the sharp time variation in $w(a)$ at more recent epochs that this particular model exhibits. Models given by the other ($m^2>0$) branch of \cref{potential} do not inherit this ambiguity to the same degree because they evolve far more linearly, and are thus not quite as sensitive to exactly which redshifts are probed by a survey.
Furthermore, the inferred values of the CPL parameters seemingly imply that this dark energy model becomes phantom in the past. However, one can clearly see that this is just an artifact of extrapolating the linear parameterization of $w(a)$ outside the range where the relevant distance observations are fit as the actual equation of state remains non-phantom across all of cosmic history (a point noted by \cite{Shlivko:2024llw} as well). Notice also that the fitted $w_0$ can differ quite substantially from the true value of the equation of state today, $w(a=1)$. 

\begin{figure*}[htbp]
    \centering
    \begin{minipage}[b]{0.9\textwidth}
        \centering
        \includegraphics[width=\textwidth]{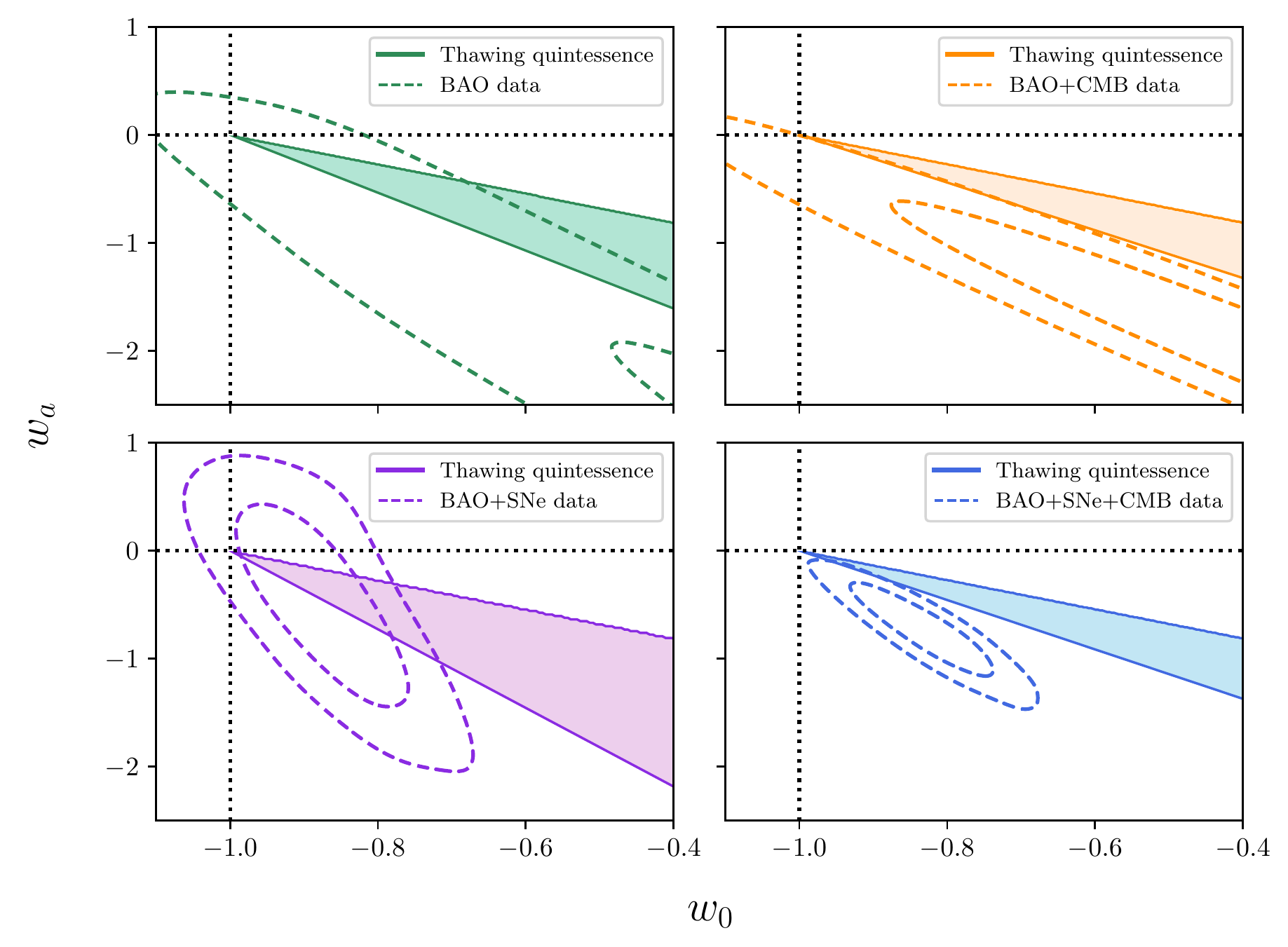}
        \label{fig:subfig1}
    \end{minipage}
    \caption{68\% and 95\% C.L. posterior distribution of the CPL parameters (dashed line) from the BAO (upper left), BAO+SNe (lower left), BAO+CMB (upper right), and BAO+CMB+SNe (lower right) data, where we use the data sets coming from  DESI data \cite{DESI:2024kob}, Pantheon+ \cite{Scolnic:2021amr}, and Planck+ACT lensing \cite{Planck:2018vyg, Planck:2019nip,ACT:2023kun, ACT:2023dou}. Overlayed are the CPL parameters for thawing quintessence, obtained fitting the surveys characteristics as explained in \cref{sec:CPLthawing}. One can see that there is reasonable overlap when only the DESI or DESI+SNe data is considered. Once the CMB is included and all of the data is taken together, thawing quintessence is barely viable.
    }\label{Fig:CPL_fit}
\end{figure*}

To identify the region of the ($w_0$, $w_a$) plane which is occupied by thawing quintessence models,
we use the Boltzmann solver \texttt{hi\_class} \cite{hi_class1,hi_class2,CLASS} to generate cosmological models with dark energy described by the potential given in \cref{potential} by sampling over $m^2$, $V_0$, $\Omega_{\rm m}$, and $h$ (note that $\varphi_{\rm ini}$ is fixed by imposing the Friedmann equation). We then fit the predictions of these models to the CPL parameterization in \cref{param} using the combined data from all of these probes, and following the procedure described in \cref{data}.  

We sample 100,000 points on a Latin hypercube in the ranges $m^2 \in [-150.0, 5]$, $V_0 \in [0.0, 2.0]$, $\Omega_{\rm m}\in [0.29, 0.33]$ and $h \in [0.65, 0.70]$. Given both $\Omega_{\rm m}$ and $h$ are highly constrained by the combined cosmological data we are using, these ranges encompass the allowed regions for these parameters. While we will fully broaden and open these priors when we proceed to the MCMC sampling, these tighter constraints will serve us for the present section. 

When it comes to the range of mass and energy scales considered, the reasoning is as follows. For $m^2>0$, as previously noted, these models converge on a universal evolution for $w(a)$ (and cosmological behavior more generally); therefore, we only need a very limited range of mass values in the positive branch to completely capture their phenomenology. Furthermore, at large $m^2>0$ these models enter the oscillatory regime, which would render them unviable candidates for dark energy. Thus, the cutoff at $m^2 = 5$. 

\begin{figure}
    \centering
    \includegraphics[width=\columnwidth]{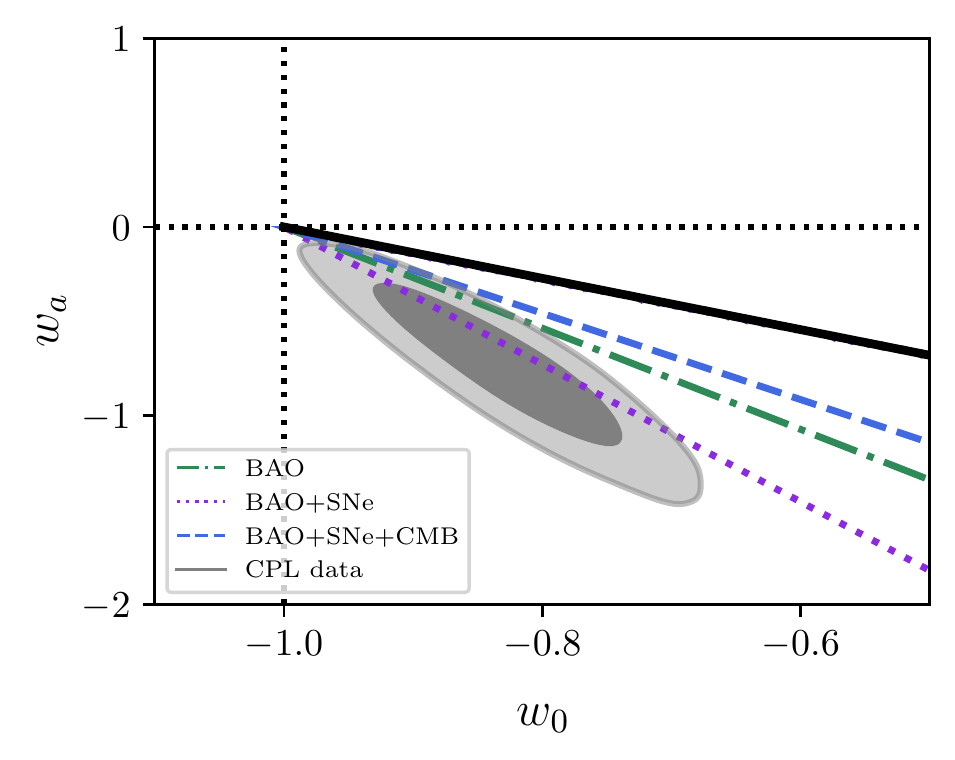}{
    \caption{Dependence of the CPL representation of thawing quintessence on survey properties. All surveys have the same upper bound (solid black line) but differing lower bounds for the corresponding ($w_0$, $w_a$) regions.}
    \label{overplot_boundaries}}
\end{figure}

The $m^2<0$ regime is, in principle, unbounded in terms of the range of masses we could look at. This is because, as mentioned above, when one increases the mass, one can always fine-tune the initial condition $\varphi_{\rm ini}$ such that the field starts closer to the top of the hill. However, as discussed in \cite{Tsujikawa:2013fta}, there is a plausible effective field theory argument for cutting off the mass range at a certain point.
We will be overly generous and consider a lower bound of $-150<m^2$ but, in practice, we will see that, given the characteristics of the surveys, we are insensitive to differences in potentials with $m^2< -{\rm few} \, \times 10$.

We are now ready to map thawing quintessence directly into the latest constraints on $w_0$ and $w_a$. We sample the dark energy model described by the potential in \cref{potential} as detailed in this section to generate the observables and then infer, for several collections of survey data, the corresponding ($w_0$, $w_a$) as described in \cref{data}. In \cref{Fig:CPL_fit}, we plot our results for a few different data choices. Here, we see how the thawing quintessence models lie in the ($w_0$, $w_a$) plane relative to the constraints derived from both the BAO measurements  and the combination of BAO and SNe measurements. As we can see, there is no evidence for deviations from $\Lambda$CDM, and there is reasonable overlap between what one might expect from thawing quintessence and the data. The situation changes substantially if we include the CMB. The CMB substantially tightens the constraints on ancillary cosmological parameters (such as $\Omega_{\rm m}$), leading to tighter constraints on ($w_0$, $w_a$). Furthermore, it probes much deeper redshifts and, as we saw in \cref{Fig:fit_model}, this pulls the region covered by thawing quintessence towards the top right hand corner of the ($w_0$, $w_a$) plane, away from the region constrained by the data.

The situation is most dramatic if we combine all the data sets together. There we find that the constraints on the ($w_0$, $w_a$) plane pull away somewhat significantly from the $\Lambda$CDM point. But we also find that the region covered by thawing quintessence models is now almost completely disjoint from the region favoured by data (albeit with a small overlap). Thus, with a combination of data sets which give the tightest constraints on ($w_0$, $w_a$) we seem to find that thawing quintessence is only marginally consistent with the data. Our results echo those found in \citet{Ye:2024ywg}. In their analysis they simplified the mapping  to ($w_0$, $w_a$) by fitting the equation of state directly; as we have shown in this paper, and argued in \cite{Wolf:2023uno}, the mapping depends on the characteristics of the survey one is considering.

In \cref{overplot_boundaries} we can see that representing thawing quintessence models in terms of ($w_0$, $w_a$) is clearly sensitive to the properties of the survey considered. They all share a common upper boundary given by the universal behavior of the $m^2>0$ models, but how far the boundary extends for the $m^2<0$ models depends on the survey properties. That is, the redshifts at which the distance measurements are taken will impact what the best fit ($w_0$, $w_a$) parameters are for a particular dark energy model, with the SNe data at recent times pulling the contour down into steeper regions of the plane, while CMB data from early times pulls the contours into the flatter regions of the plane.

\begin{figure}
    \centering
    \includegraphics[width=\columnwidth]{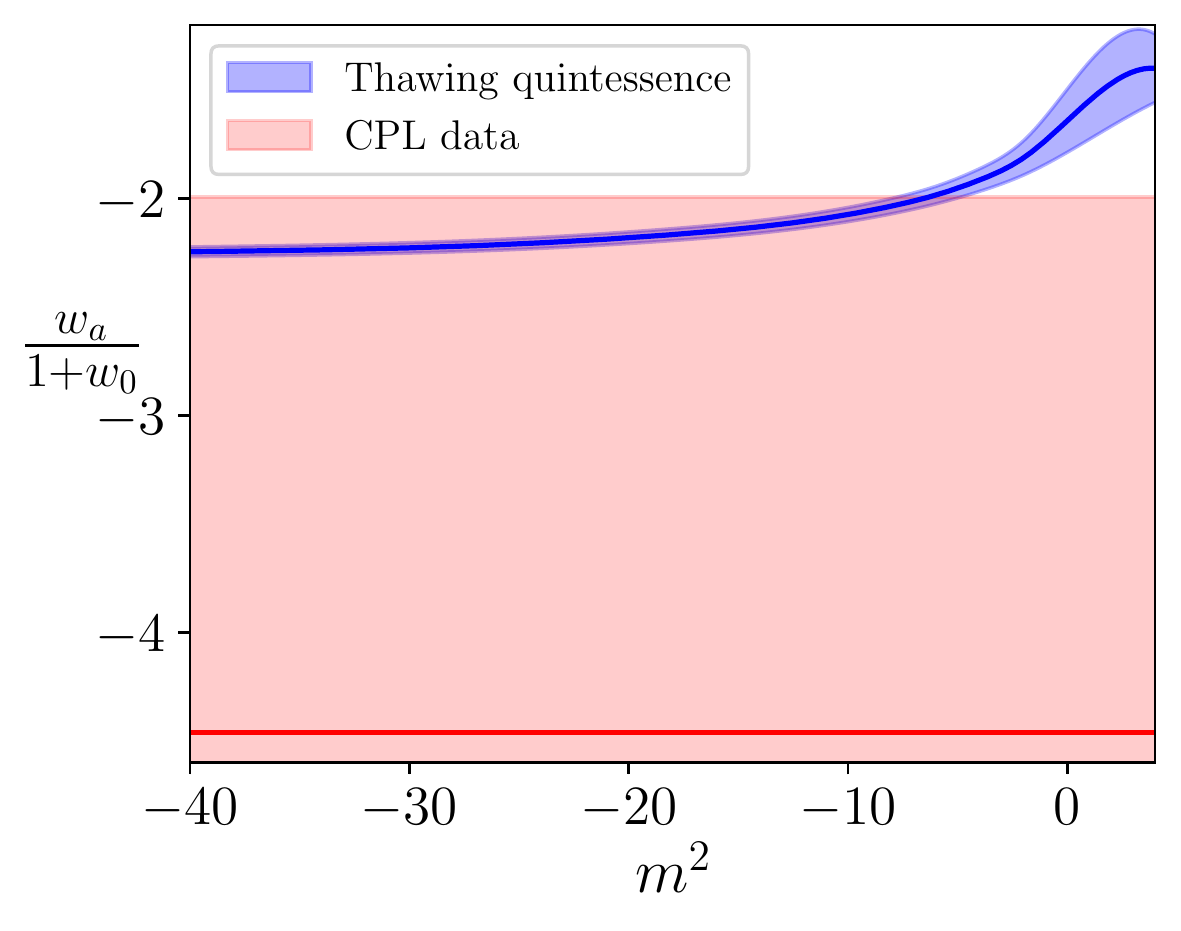}
    \caption{Blue: The predicted slope $ w_a/(1+w_0)$ of thawing quintessence models in the ($w_0$, $w_a$) plane for survey characteristics of BAO+CMB+SNe data. Red: CPL parameterization data constraints (as \cref{fig: DESI contour}) . The shaded areas represent $2\sigma$ regions for both thawing quintessence and the data.}
    \label{Fig:sigma_mass}
\end{figure}

It is useful to understand the role of $m^2$ in the mapping between thawing quintessence and the ($w_0$, $w_a$) plane. We do so by focusing on the slope of quintessence models in the ($w_0$, $w_a$) plane, which we define to be $ \alpha = w_a/(1+w_0)$. For a given survey, this slope is primarily set by $m^2$, with a small variance due to the other parameters. 
In \cref{Fig:sigma_mass}, we plot the slope as a function of $m^2$ for the combination of data which includes BAOs, SNe and the CMB, following the procedure of \cref{data}. We can see that as $m^2$ becomes increasingly negative, the slope tends towards a constant, $\alpha \simeq -2.25$ and the lower $m^2$ models all eventually become indistinguishable from each other. We can compress the contours from \cref{fig: DESI contour} into the red horizontal band in \cref{Fig:sigma_mass}. Again, as we saw in the bottom right hand panel of \cref{Fig:CPL_fit}, through the lens of the CPL parametrization, thawing quintessence is only marginally consistent with current data.

In summary, the CPL parametrization is a somewhat flawed form of data compression, but it is useful to gain an understanding of how a particular model compares to constraints from data. By finding the $(w_0, w_a)$ for thawing quintessence (for particular survey specifications) we are, to some extent, determining the prior distribution of this class of theories in the CPL parameter plane. We can then compare it to the likelihood for $(w_0, w_a)$ derived from the data and see if there is significant or very little overlap. We see that, for the case in which we use a combination of BAO, SNe and CMB data there is, indeed, very little overlap between the thawing quintessence prior and the likelihood.

\section{Constraints on thawing quintessence}\label{Sec:MCMCresults}

So far we have seen that thawing quintessence is only marginally consistent with a compilation of current data through the lens of the CPL parametrization. But one {can} look at the problem in a different way and take thawing quintessence as the unique cause for late time acceleration and constrain its parameters. This is tantamount to considering a prior on dark energy which is that {of} thawing quintessence. Thus, instead of constraining ($w_0, w_a$) from the data and then comparing these constraints to what is predicted by thawing quintessence for those parameters, one constrains the thawing quintessence parameters directly.

The cosmological model one is now constraining is effectively described in terms of a set of parameters. ($V_0$, $m^2$, $\Omega_{\rm m}$, $\ldots$) and, in principle, one would like to assume uninformative, uniform priors on all these parameters. This is particularly true of $m^2$ as this is the parameter that is responsible for time evolution in the equation of state. One needs to be sure that the prior is such that it does not {artificially} enhance or suppress evidence for non-zero $m^2$.

\begin{figure}
    \centering
    \includegraphics[width=\columnwidth]{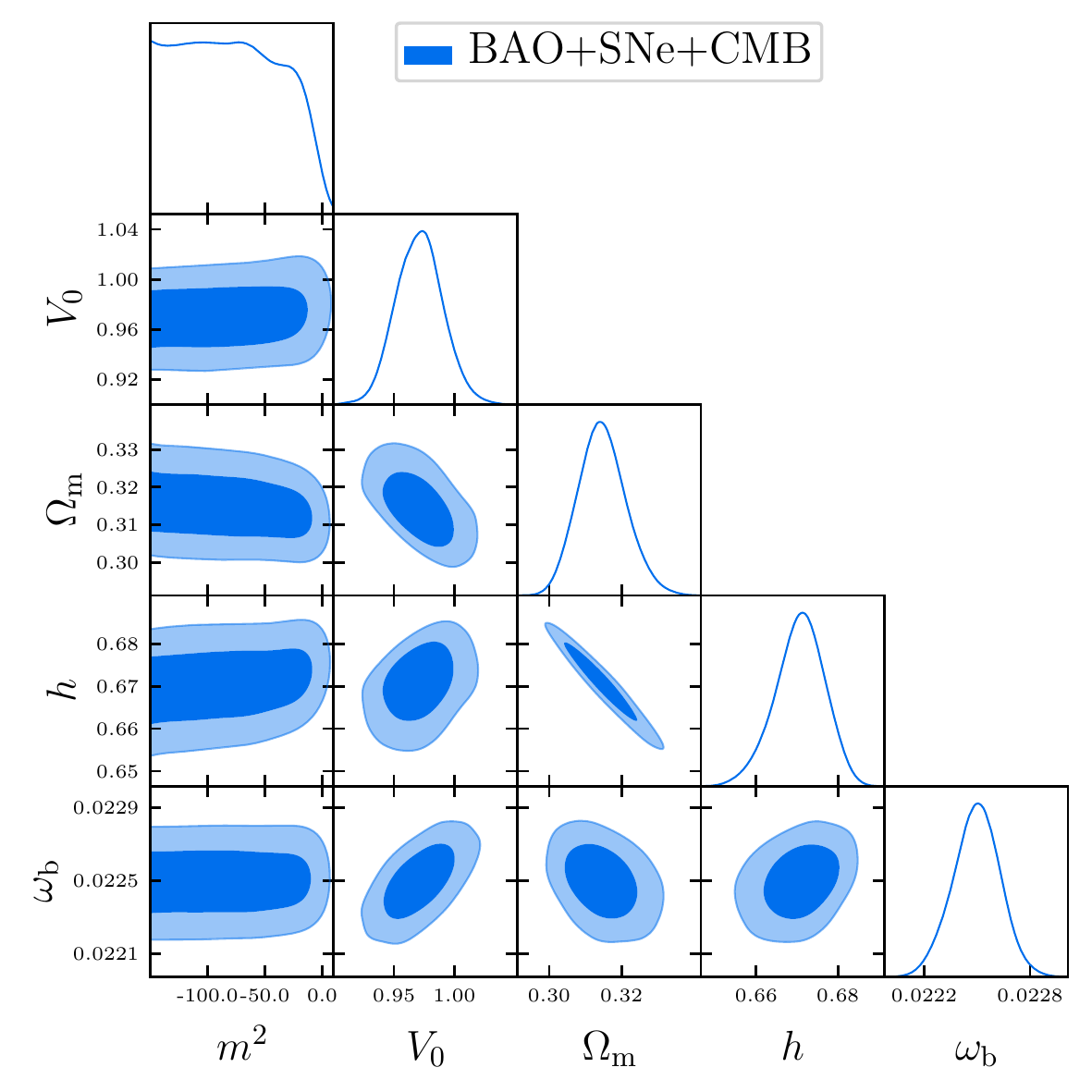}
    \caption{68\% and 95\% C.L. posterior distributions for thawing quintessence from the combination of DESI BAO data \cite{DESI:2024mwx}, Pantheon+ SNe data \cite{Scolnic:2021amr}, and CMB data \cite{Planck:2018vyg, Planck:2019nip,ACT:2023kun, ACT:2023dou} obtained from sampling uniformly in the parameters given by Table~\ref{priors}. See Fig.~\ref{Fig:mcmc_m2_full}, in \cref{sec.fullposterior}, the full corner plot with all cosmological parameters.}
    \label{fig:mcmc_m2}
\end{figure}

\begin{table}[h!]
\centering
\begin{tabular}{|c|c|}
\hline
\multicolumn{2}{|c|}{\textbf{Cosmological Parameters}} \\
\hline
$\omega_b$ & $\mathcal{U}[0.005, 0.1]$ \\
$\Omega_m$ & $\mathcal{U}[0.01, 0.99]$ \\
$H_0 \ [\text{km/s/Mpc}]$ & $\mathcal{U}[20, 100]$ \\
$n_s$ & $\mathcal{U}[0.8, 1.2]$ \\
$\ln 10^{10} A_s$ & $\mathcal{U}[1.61, 3.91]$ \\
$\tau$ & $\mathcal{U}[0.01, 0.8]$ \\
\hline
\multicolumn{2}{|c|}{\textbf{Quintessence Parameters}} \\
\hline
$m^2$ & $\mathcal{U}[-150, 10]$ \\
$V_0$ & $\mathcal{U}[-2.0, 2.0]$ \\
\hline
\end{tabular}
\caption{Prior distributions used in the cosmological parameter inference. $\mathcal{U}(a, b)$ stands for an uniform distribution in the range $[a, b]$. For Planck's nuisance parameters, we use the \texttt{Cobaya} default values, specified in  \cite{Planck:2019nip}.}\label{priors}
\end{table}

There are a few aspects of the analysis which need to be highlighted. First of all, we saw that, as $m^2$ becomes ever more negative, the more indistinguishable $w_a/(w_0+1)$ {becomes} between models of different mass. This can be understood from the dynamics of the scalar field described above -- for large, negative values of $m^2$, the scalar field evolves very rapidly but only at very late times. So, from the point of view of the redshift ranges being covered by the surveys which may not probe such late times, the cosmological evolution of the thawing quintessence models will become indistinguishable. Indeed, we shall see that, for a fixed compilation of data, the likelihood of the thawing quintessence model will tend to a constant as $m^2\rightarrow -\infty$. 

We have already highlighted the fact there should be a degeneracy for $m^2>0$ which, in turn, may lead to an enhanced prior for positive masses. This could, potentially, be a source of concern in interpreting the posterior constraints we obtain on $m^2$ as ``volume effects'' could possibly bias the constraints towards $m^2\ge 0$. As we will see, that is not the case -- the likelihood more than compensates for the effect of the prior there and the posterior favors $m^2<0$.

\begin{figure}
    \centering
    \includegraphics[width=\columnwidth]{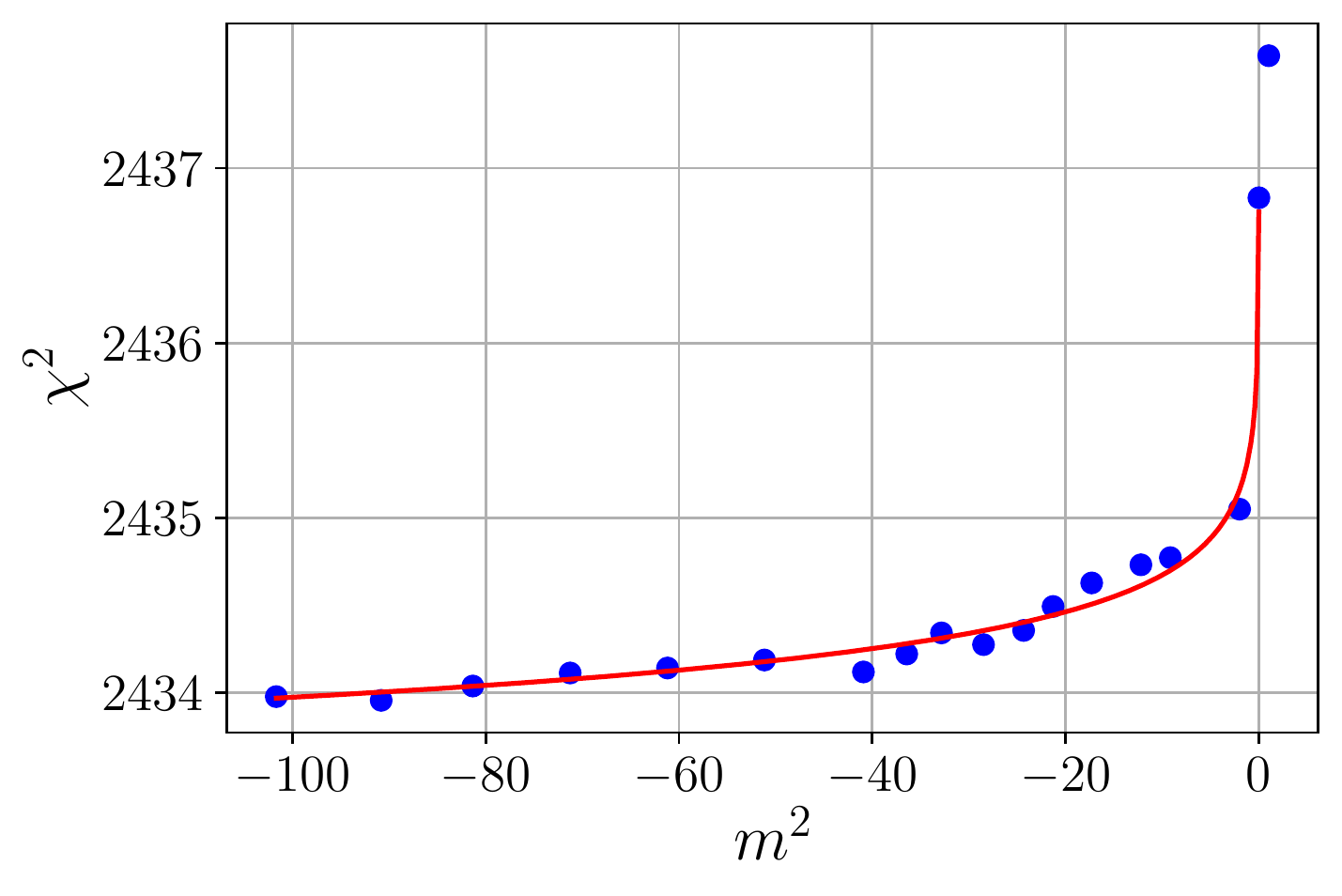}
    \caption{Minimum $\chi^2=-2\ln {\cal L}$ (where ${\cal L}$ is the likelihood) as a function of $m^2$. Blue dots are the result of the numerical minimization while the red line is the best fit to the dots for $m^2<0$, shown for illustrative purposes.
    }
    \label{chi2 vs m2 plot}
\end{figure}

A crucial part of our analysis is ensuring that, for $m^2<0$, the prior is flat. This is non-trivial given how one generates the observables with \texttt{hi\_class}. Given a value of $m^2$, the Boltzmann-solver has to iterate to find the corresponding initial condition, $\varphi_{\rm ini}$, which will lead to the cosmology with the correct values of the remaining cosmological parameters. This is a time consuming procedure for each value of $m^2$ and highly dependent on a ``good'' choice for the iterative procedure.

To speed up the iterative search, we have implemented a novel procedure, described in \cref{sec.symbolic}: we have used Symbolic Regression to determine an analytic expression for $\varphi_{\rm ini}$ as a function of the remaining parameters, most notably $m^2$, which is accurate to within $1\%$. This is then used as the initial guess for the solver and allows us to efficiently generate a flat prior for $m^2$. We confirm that our procedure leads to flat priors by running the inference without any data (in practice, by setting the likelihood to a constant) as can be see in \cref{fig: m2 prior} in \cref{sec.m2}.

With the priors correctly under control, we proceed to infer the posterior on $m^2$ for a range of data sets. 
{
We use the same data as in the DESI analysis \cite{DESI:2024mwx}; i.e., DESI 2024 BAO measurements \cite{DESI:2024mwx, DESI:2024uvr, DESI:2024lzq}, the SNe Ia Pantheon+ samples \cite{Brout:2022vxf, Scolnic:2021amr}, the TTTEEE CMB measurements from Planck 2018 \cite{Planck:2019nip, Planck:2018vyg},  and the CMB lensing from ACT DR6 \cite{ACT:2023dou, ACT:2023kun}. 

DESI 2024 BAO measurements consist of a set of six measurements of the angular diameter distance, $D_M$, the Hubble parameter, $D_H = c/H$, and the angle-averaged quantity $D_V = (z D_M^2 D_H)^3$, relative to the sound horizon, obtained from measurements of the BAO scale, using 6 million galaxies in the range of redshifts $0.1 < z < 4.2$.

The Pantheon+ sample is made of 1701 light curves from 1550 different SNe Ia ranging $z = 0.001$ to $2.26$, obtained with observations of DES \cite{1811.02378, 1811.09565}, Foundation \cite{1711.02474}, Pan-STARRS \cite{1710.00845}, Supernova Legacy Survey \cite{1401.4064}, SDSS \cite{1107.5106}, HST \cite{astro-ph/9903229, astro-ph/0104455, astro-ph/0402512, astro-ph/0611572, 1105.3470, Riess:2017lxs} and multiple low-$z$ samples (see \cite{Scolnic:2021amr} for the full list).

From Planck, we use the auto- and cross-correlation of the CMB temperature and $E$-mode polarization fields. We use the low-$\ell$ power spectra, obtained with the ``Commander'' component separation algorithm, in the range $2 \leq \ell \leq 29$ and the high-$\ell$ \texttt{plik} likelihood covering $30 \leq \ell \leq 2508$ for the temperature auto-correlation ($TT$) and $30 \leq \ell \leq 1996$ for the $TE$ and $EE$ components \cite{Aghanim:2019ame}. We marginalize over the nuisance parameters related to the foregrounds calibration.

Finally, we use CMB lensing from ACT DR6 as provided by the official likelihood\footnote{\url{https://github.com/ACTCollaboration/act\_dr6\_lenslike}} with the variant \texttt{act\_baseline}\cite{ACT:2023dou, ACT:2023kun}, consisting of the ACT CMB reconstructed lensing power spectra between the scales $40 < L < 763$. 

We use the likelihoods as implemented in \texttt{Cobaya} \cite{Cobaya, Cobaya2},
which we then use to explore the parameter space using a Metropolis-Hasting
algorithm \cite{metropolismc, hastingsmc, Lewis:2002ah, Lewis:2013hha} by
sampling the parameters in Table~\ref{priors} and imposing uniform priors. We
address convergence using the Gelman-Rubin criterion \cite{gelmanrubin},
imposing that $R-1 < 0.02$ in the diagonalized parameter space. Finally, we
compute the theory predictions with \texttt{hi\_class} and model the non-linear
matter power spectrum with the \textsc{halofit} algorithm \cite{halofit1,
halofit2}\footnote{For efficiency, we do not follow the recommended set up for
ACT~\cite{Hill:2021yec}. We have checked that the
impact is negligible, with $\Delta \chi^2 \sim 0.2$ for the best fit model of 
the full data set.}.
}

In \cref{fig:mcmc_m2}, we show the posteriors for the cosmological parameters, including ($V_0$, $m^2$) for the combination of BAOs, SN and CMB data sets. The posterior for $m^2$ is qualitatively, as expected, favoring $m^2<0$ and flat out to the minimum (most negative) $m^2$ considered. Indeed, the data favors a thawing quintessence model over the $m^2=0$, $\Lambda$CDM model. Unfortunately, given the improper nature of the prior on $m^2$ (i.e., it is in principle, constant for $m^2<0$ and unbounded from below), it is dependent on our choice of the lower bound, and thus impossible to make a clear (prior independent) statement on how much we can rule out $\Lambda$CDM vis-\`a-vis thawing quintessence. 
This situation is not too dissimilar to what one find when constraining Jordan-Brans-Dicke (JBD) models \cite{Joudaki:2020shz}, where one can show how constraints on the JBD parameter which characterizes the non-minimal coupling of the scalar field to gravity is completely dependent on the choice of priors.

It is useful, nevertheless, to try and assess if the data has a preference for thawing quintessence relative to $\Lambda$CDM. The simplest approach is to look at the $\chi^2=-2\ln {\hat{\cal L}}(m^2)$, where ${\hat {\cal L}}$ is the likelihood of the model with mass $m^2$ and maximized over the other parameters\footnote{Note that, for speed, we use the \textit{lite} version of Planck likelihood in the minimization.}. From Fig.~\ref{chi2 vs m2 plot}, where we plot the $\chi^2$ of the model as a function of $m^2$, we can see that the $\chi^2$ assymptotes to $\chi^2_f\simeq 2434$ for sufficiently negative values of $m^2$. If we now compare the model with $m^2=0$ (where there is no thawing) with one with a suitably negative value of $m^2$, we find a $\Delta \chi^2 \simeq 2.8$ showing a very marginal preference for thawing quintessence.

We can further refine the comparison by considering the Akaike Information Criterion (AIC) and the Bayesian Information Criterion (BIC) as these incorporate more information about the model \cite{Liddle:2007fy, AIC_1974, BIC_1978}. The AIC is given by
$
 {\rm AIC}=2k+\chi^2    
$,
where $k$ is the number of parameters in the model. We can now compare the $\Lambda$CDM model with a thawing model with negative $m^2$. Apart from the cosmological parameters,
$\Lambda$CDM has $\{V_0\}$, whereas thawing quintessence has $\{V_0, m^2, \varphi_{\rm ini}\}$. In the case of thawing quintessence, $\varphi_{\rm ini}$ (or any of the others) is fixed by the Friedmann equation. In the case of $\Lambda$CDM, this is also true: $V_0$ is fixed by the Friedmann equation. As a consequence,  $\Delta k \equiv k^{\rm \Lambda CDM} - k^{\rm Thawing} = -2$ and $\Delta {\rm AIC}\simeq -1.2$, rendering them effectively (statistically) indistinguishable, albeit a mild preference towards $\Lambda$CDM. If we consider the BIC, which is given by
$
    {\rm BIC}=k\ln(n)+\chi^2 
$,
where $n\simeq 2400$ is the number of data points, we have that $\Delta {\rm BIC}\simeq -12.8$ which strongly favours the $\Lambda$CDM model. Thus, we can conclude that there is little evidence for thawing quintessence with current cosmological data.

We can compare these results with what we find if, instead, we look at the CPL parametrization. There is a more marked difference between the best fit $(w_0,w_a)$ model as compared to $\Lambda$CDM. There we find that
$\Delta\chi^2\simeq 6.8$, $\Delta {\rm AIC}\simeq 2.8$ and $\Delta {\rm BIC}\simeq -8.8$ indicating that, from the point of view of two of our measures, the best fit CPL model is slightly preferred while, again, from the point of view of the ${\rm BIC}$, it is not.

\begin{figure}
    \centering
    \includegraphics[width=\columnwidth]{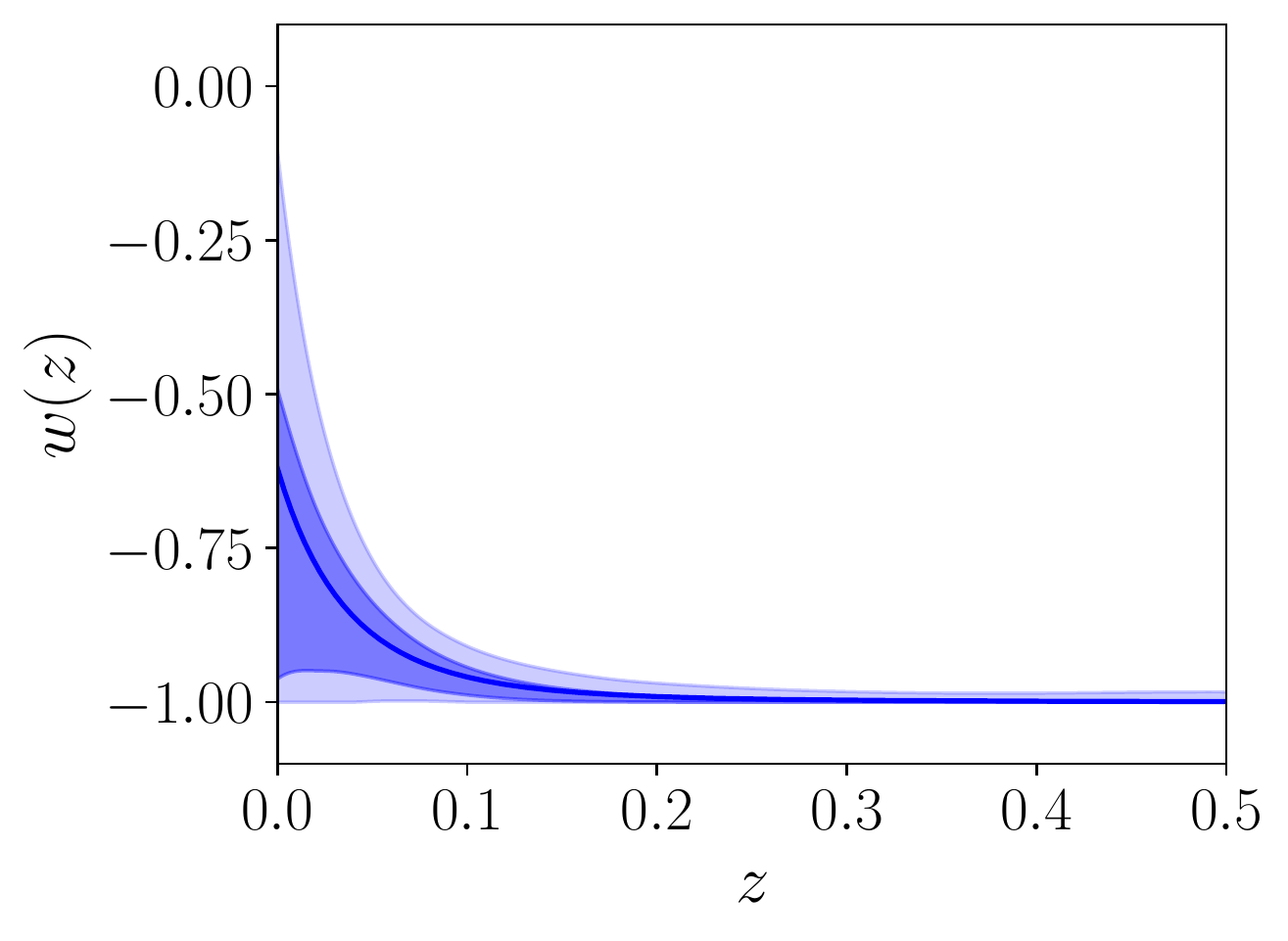}
    \caption{Reconstruction of the equation of state (1- and 2-$\sigma$ levels) from the combination of DESI BAO data \cite{DESI:2024kob}, Pantheon+ SNe data \cite{Scolnic:2021amr}, and CMB data \cite{Planck:2018vyg, Planck:2019nip,ACT:2023kun, ACT:2023dou}. The equation of state is indistinguishable from a cosmological constant for most of the Universe's history, becoming unconstrained at recent times. 
    }
    \label{fig:wzconstraint}
\end{figure}

\begin{figure}
    \centering
    \includegraphics[width=\columnwidth]{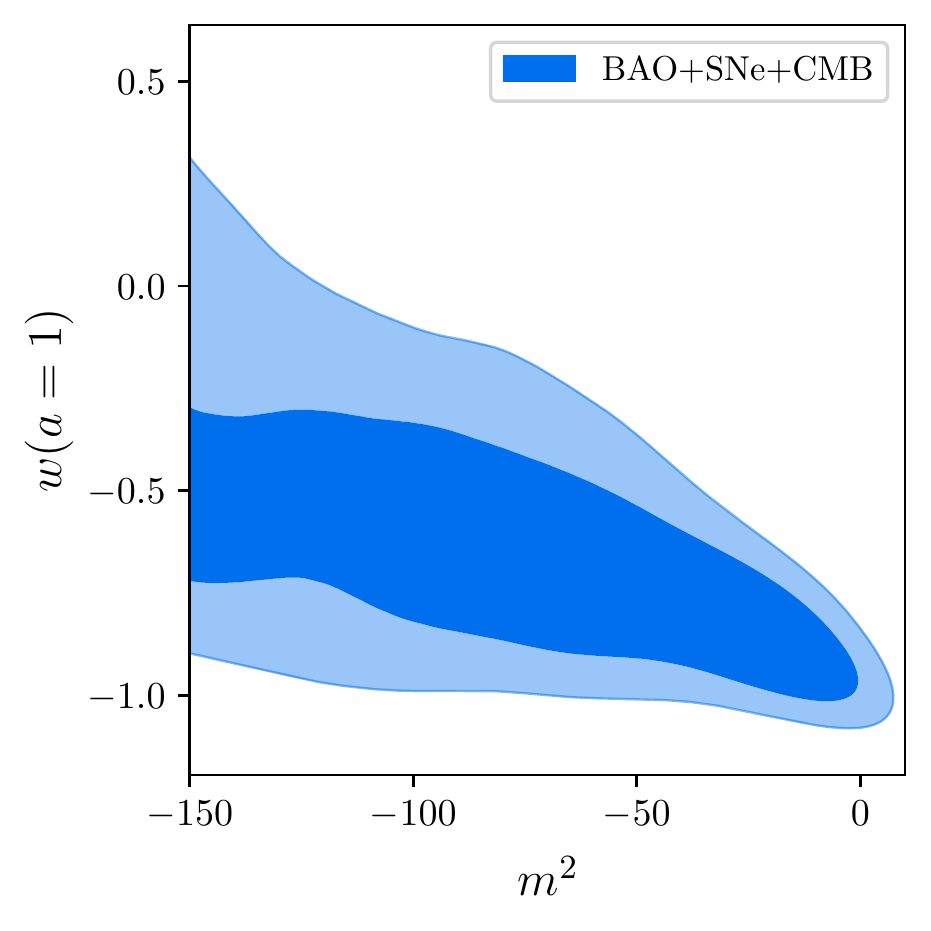}
    \caption{68\% and 95\% C.L. posterior distribution of $w(a=1)$ (not to be confused with the CPL parameter $w_0$) and $m^2$ in thawing quintessence from the combination of DESI BAO data \cite{DESI:2024kob}, Pantheon+ SNe data \cite{Scolnic:2021amr}, and CMB data \cite{Planck:2018vyg, Planck:2019nip,ACT:2023kun, ACT:2023dou}.}
    \label{fig:w0constraint}
\end{figure}

Finally, we take a look again at the equation of state for thawing quintessence, given the current data. In \cref{fig:wzconstraint}, we show constraints on the value of $w(z)$ across the recent expansion history going out to $z\simeq0.5$ assuming that dark energy is described by the present model of thawing quintessence. This shows that, until quite recently, if dark energy is described by thawing quintessence, $w(a)$ would be indistinguishable from a cosmological constant, 
before becoming quite unconstrained at recent times. Even here though, from the perspective of constraining directly in terms of ($V_0$, $m^2$), a value for the equation of state indistinguishable from $\Lambda$CDM is still allowed.
It is also interesting to see, in \cref{fig:w0constraint}, constraints on the value of the equation of state $w(a=1)$ today in terms of $m^2$. Unsurprisingly, the larger the negative hilltop mass becomes, the more unconstrained this value becomes as the equation of state can evolve very fast at these masses. 

\section{Conclusion}\label{sec:conclusion}

In this paper we have looked at the evidence for evolving dark energy in the form of quintessence, i.e., a scalar which is minimally coupled to gravity and has a canonical kinetic energy. Current constraints point, if anything, towards thawing quintessence, where the scalar field starts to evolve at late times. Following \cite{Wolf:2023uno}, we consider the potential which encompasses all possible thawing quintessence-like behaviour. 

First, we transform its equation of state $w(a)$ into the CPL parametrization, ($w_0$, $w_a$), and compare its constraints with current data. We find that the evidence for thawing quintessence, from the point of view of CPL, is marginal at best, in agreement with what was found in \citet{Ye:2024ywg}. An essential aspect of our analysis is that we determine the ($w_0$, $w_a$) parameters of the quintessence model using exactly the same procedure as was used to obtain them from the data. This means we are comparing like with like and highlights the fact that a given quintessence model does not map on to a unique value of  ($w_0$, $w_a$).

We then directly constrain the parameters of thawing quintessence, ($V_0$, $m^2$), and find there is some evidence for evolution, distinguishing it from a cosmological constant. The posterior distribution of $m^2$ is such that it is difficult to quantify how strong the evidence is. Nevertheless, by looking at the relative likelihoods of the best fit thawing quintessence model relative to $\Lambda$CDM, and using other information criteria, we find that thawing quintessence is between mildly ($\Delta {\rm AIC}\simeq -1.2$) and strongly ($\Delta {\rm BIC}\simeq -12.8$) disfavored with respect to $\Lambda$CDM. In other words, there is no preference for thawing quintessence.

As we have argued, the quadratic potential we are considering here covers, in our view, a vast range of thawing quintessence models. As shown in \citet{Wolf:2023uno}, this means that we can span a large wedge in the $(w(a=1),-w'(a=1)$) parameters. But this is not the same as the ($w_0, w_a$) plane, as explained above. One might, however, consider thawing like equations of state which deviate from the ones considered here and that could have a more significant overlap with the ($w_0, w_a$) constraints. Qualitatively they would have to behave as follows: the equation of state would start off at $-1$, then, at some intermediate redshift (coincident with the redshift ranges of the data we are considering here) it would thaw rapidly to then stabilize again at some constant value until it reaches today. Even so, it would be difficult to obtain the slopes detected in the ($w_0, w_a$) given the role of the CMB data (see Fig \ref{Fig:fit_model}). Furthermore, it would require some creative engineering of a potential that would lead to that behaviour.

The fact that thawing quintessence does so poorly relative to the view of CPL parameterization for dark energy is intriguing. Does it mean we need to consider a broader family of scalar field models, with non-minimal couplings or non-canonical kinetic energy? This might be the case, although as argued above and by others, the evidence for phantom evolution from ($w_0$, $w_a$) is misleading. On the other hand, \citet{Ye:2024ywg} have come up with a proof of concept that it is possible to construct a (simple) well behaved non-minimally coupled model that leads to phantom behaviour and is consistent with the data. Dubbed ``Thawing Gravity'' it will be interesting to see if it is consistent with other constraints on fifth forces which arise in non-minimally coupled theories. At the very least, it raises the possibility that we are detecting evidence of non-minimal coupling on cosmological scales.

As stated in the introduction, the CPL model is a blunt instrument and clearly too blunt to make any strong inferences about the underlying microphysics (or even just the qualitative behavior) of dark energy. Its strength comes from the fact that it is very simple as it only depends on two parameters. Other proposals have been put forward which are as minimal but run into the same problems as the CPL parameters. On the other hand, there are a number of more complicated parameterizations which track the evolution of dark energy more accurately, but these run the risk of overfitting and being plagued by degeneracies between the parameters. 
Clearly, an all purpose, simple yet accurate parameterization is still lacking.

The analysis in this paper hinges on the data being correct; this might well not be the case as pointed in \cite{Efstathiou:2024dvn,Efstathiou_2024}. It would be wise to reserve judgement until there are more, and better, constraints on the expansion history of the Universe. We expect this to be the case over the coming years.

\section*{Acknowledgements}

We acknowledge the \texttt{hi\_class} developers,
Emilio Bellini, Miguel Zumalac\'arregui and Ignacy Sawicki, for sharing a private version of the code.
We thank David Alonso, Harry Desmond, Constantinos Skordis and Maria Vincenzi for useful discussions. WW is supported by St.~Cross College, Oxford. CGG is supported by the Beecroft Trust.
DJB is supported by the Simons Collaboration on ``Learning the Universe.'' PGF is supported by STFC and the Beecroft Trust.

{\it Software}:  We made extensive use of {\tt hi\_class}\cite{hi_class1,hi_class2,CLASS}, {\tt Cobaya} \cite{Cobaya, Cobaya2} and the {\tt numpy} \cite{oliphant2006guide, van2011numpy}, {\tt scipy} \cite{2020SciPy-NMeth}, {\tt matplotlib} \cite{Hunter:2007}, {\tt GetDist} \cite{Lewis:2019xzd} and {\tt pyoperon} \cite{Burlacu_2020} python packages. 

For the purposes of open access, the authors have applied a Creative Commons Attribution (CC BY) licence to any Author Accepted Manuscript version arising.

\appendix

\section{Compressing the data.}\label{data}

{
To account for the surveys constraining power when estimating $(w_0, w_a)$, for simplicity, we compress the CMB and SNe samples into cosmological background quantities that we can then fit following \cref{sec:CPLthawing}. 
}

\subsection{CMB}

\begin{figure}
    \centering
    \includegraphics[width=\columnwidth]{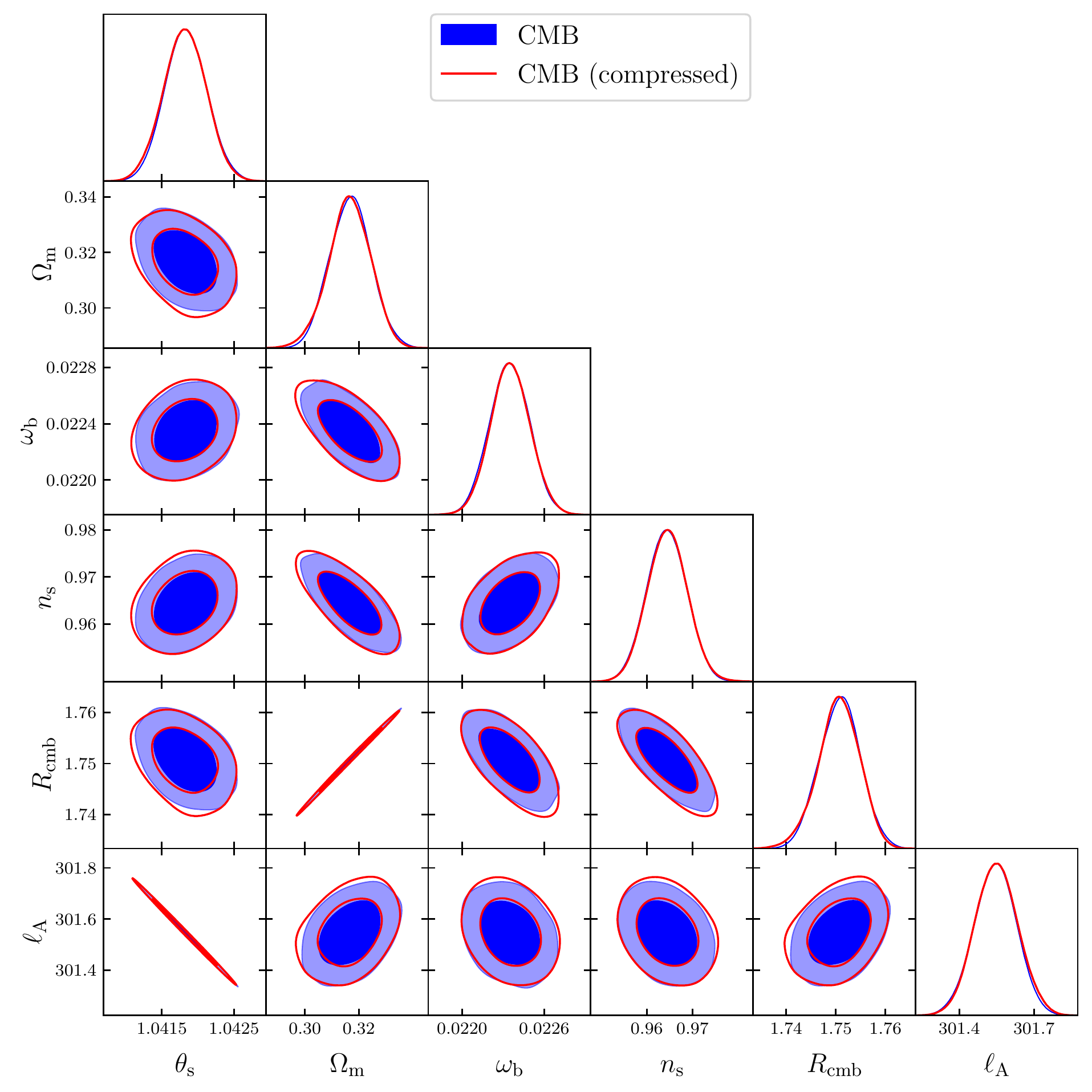}
    \caption{Posterior distribution (68\% and 95\% C.L.) obtained with the full CMB likelihood (blue) in comparison with the compressed version (red) for the $\Lambda$CDM model. The compressed CMB data recovers the posterior.
    }
    \label{fig:lcdm_compressed}
\end{figure}

{
We compress the CMB data (Planck's TTTEEE \cite{Planck:2018vyg, Planck:2019nip} and ACTDR6 lensing \cite{ACT:2023dou, ACT:2023kun}) into the background quantities \cite{Komatsu_2009, Planck:2015bue, Chen:2018dbv}: the angular scale of the sound horizon 
\begin{equation}
    \ell_{\mathrm{A}} = \pi \frac{D_A(z_*)}{r_s(z_*)} = \frac{\pi}{\theta_*},
\end{equation}
and the CMB shift parameter
\begin{equation}
    R(z_{*}) = \sqrt{\Omega_{\rm m} H_0^2} D_A(z_*),
\end{equation}
at last scattering (i.e., the photon decoupling redshift $z_{*}$). In addition, one also needs to add constraints on $\omega_b = \Omega_b h^2$ and $n_s$ to recover the full likelihoods constraints. We show in \cref{fig:lcdm_compressed} that a Gaussian likelihood using these four parameters, with mean and covariance shown in Table~\ref{tab:SNe_CMB_compressed}, is good enough to reproduce the CMB posterior distributions. 
}

\subsection{SNe IA}

\begin{figure}
    \centering
    \includegraphics[width=\columnwidth]{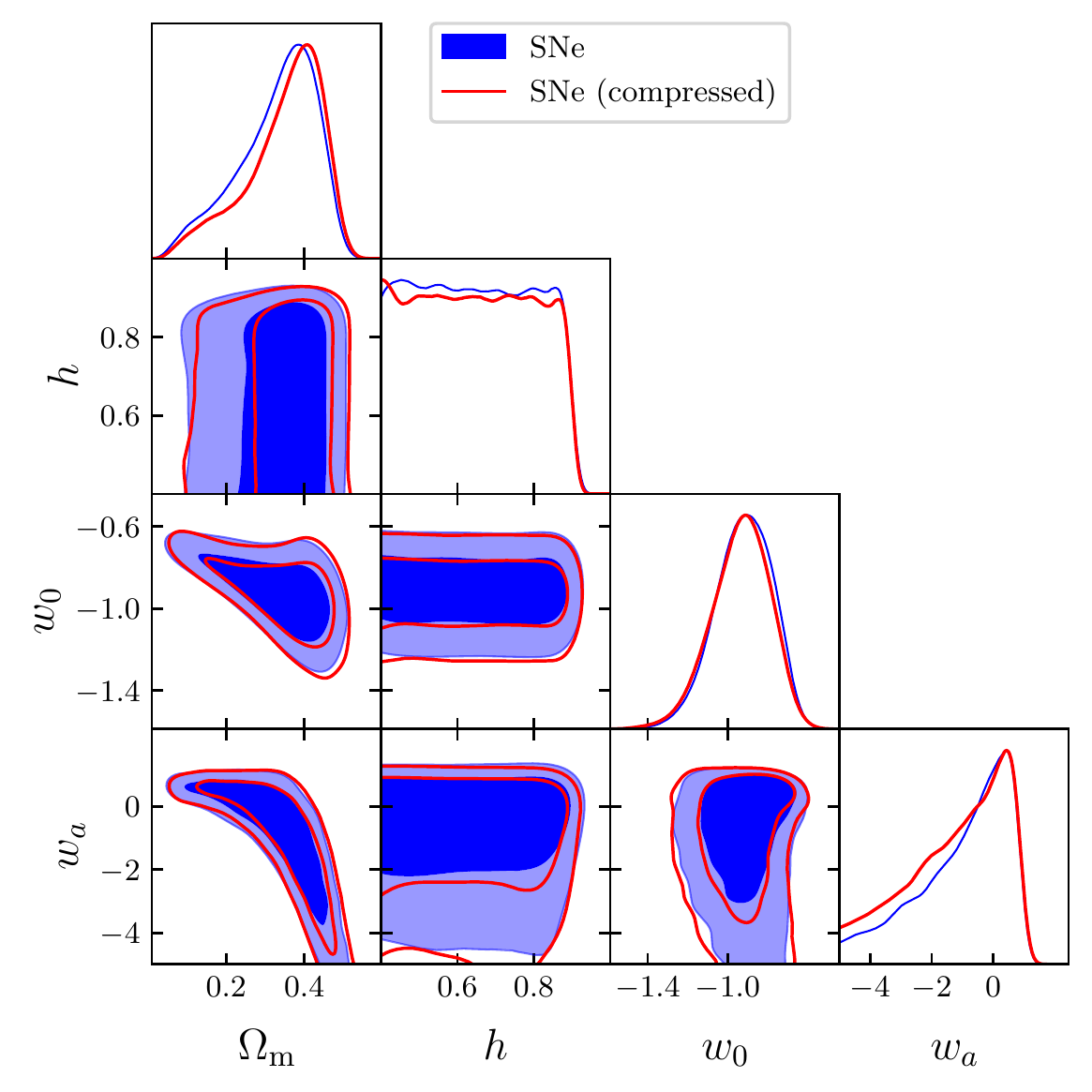}
    \caption{Posterior distribution (68\% and 95\% C.L.) obtained with the Pantheon+ SNe Ia likelihoods and its compressed version. One can see the later agrees well with the full likelihood.
    }
    \label{fig:w0wa_SNe_compressed}
\end{figure}

{
We compress the SNe Ia data following \cite{Riess:2017lxs}. Briefly, the SNe data measures 
\begin{equation}
    D_L = \frac{1+z}{H_0} \int \frac{\md z}{E(z)},
\end{equation}
where we have defined the expansion rate $E(z) \equiv H(z) / H_0$. One can compress the SNe data in model independent measurements of $1/E(z)$: instead of obtaining $1/E(z_i)$ from a cosmological model we can directly compute it as a spline: $1/E(z) = {\rm spline}(1/E(z_i))$, with nodes $1/E(z_i)$. We use a cubic spline with nodes defined at $z = \{0.07, 0.2, 0.35, 0.55, 0.9, 1.5, 2, 2.5\}$, which are the same as in \cite{Riess:2017lxs}, extended to higher redshift due to the extra SNe in the Pantheon+ sample. Pantheon+ is not able to constrain the highest redshift, so we drop that node from the compressed data set. Furthermore, we approximate the compressed likelihood as a Gaussian distribution with the mean and covariance shown in \cref{tab:SNe_CMB_compressed}. The theory code to compress the data and the \texttt{Cobaya} likelihood can be found in \url{https://gitlab.com/carlosggarcia/mcmc-likelihoods}. Finally, we show in \cref{fig:w0wa_SNe_compressed} that we can accurately recover the posterior distribution with the compressed likelihood.
}

\begin{figure}
    \centering
    \includegraphics[width=\columnwidth]{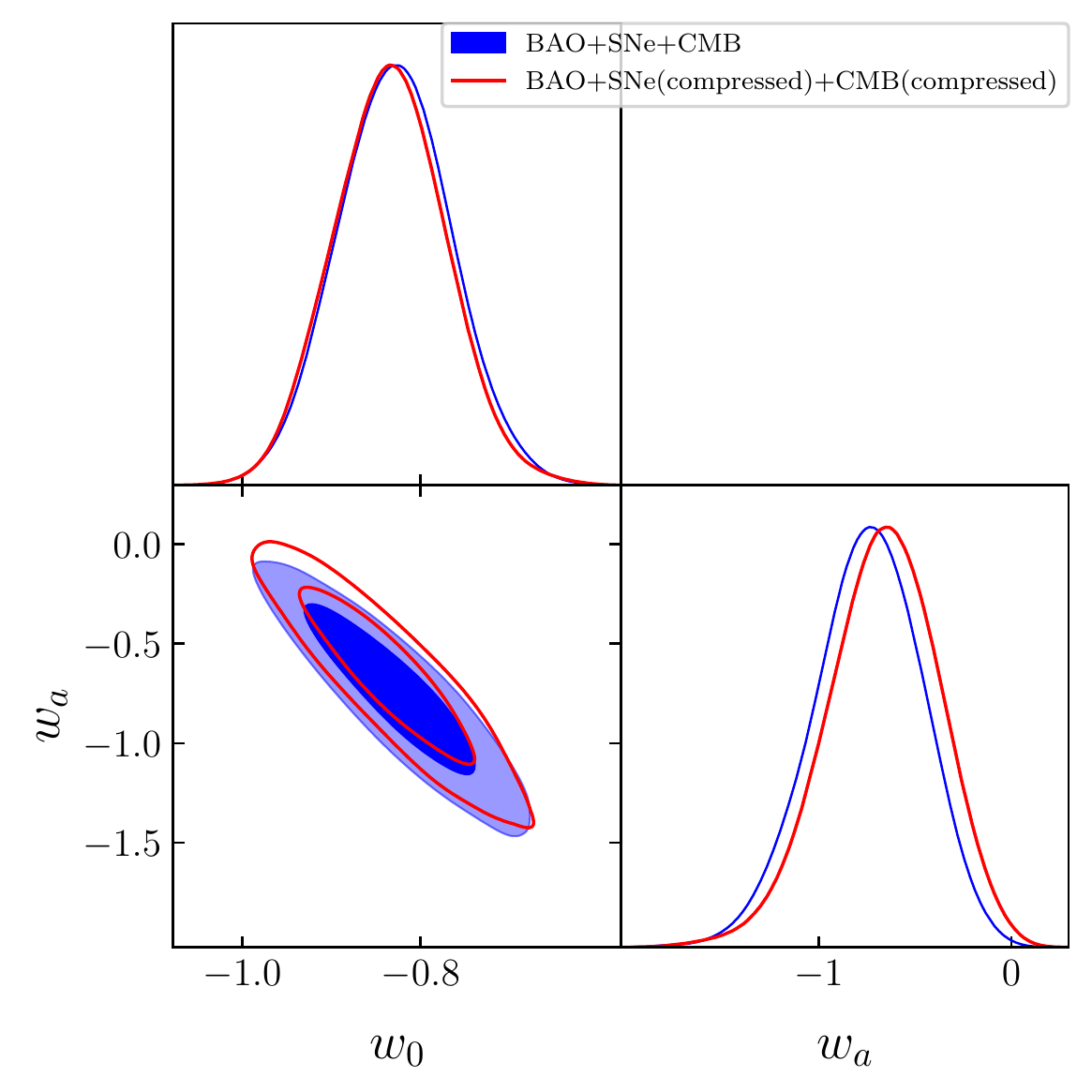}
    \caption{Posterior distribution (68\% and 95\% C.L.) obtained with DESI BAO and the full CMB and Pantheon+ SNe likelihoods (blue), compared to that obtained with DESI BAO and the compressed likelihoods of the CMB and SNe. The compressed data, combined with DESI, can clearly recover the constraints on ($w_0$, $w_a$) that come from the full complement of data to a high degree of accuracy.}
    \label{fig: w0wa compressed}
\end{figure}

Combining the results of our data compression for both the CMB and SNe IA, we show in Fig.~\ref{fig: w0wa compressed} that the compressed data can successfully recover the headline results from \cite{DESI:2024mwx} and depicted in Fig.~\ref{fig: DESI contour}. That is, when combined with DESI BAO data, the compressed CMB and SNe IA variables reproduce the results of the full data to a very good approximation.

\begin{center}
\begin{table*}
    \centering
    \begin{tabular}{|c|c|ccccccc|}
    \hline
    \multicolumn{9}{|c|}{Pantheon+ SNe compressed likelihood}\\
    \hline
    Redshift & Mean $1/E(z)$ $\pm$ error & \multicolumn{7}{c|}{Correlation matrix}\\
    \hline
    0.07 & $0.947 \pm 0.014$ & 1.00 & 0.15 & 0.59 & 0.10 & 0.15 & -0.05 & 0.07 \\
    \hline
    0.20 & $0.8982 \pm 0.0095$ & 0.15 & 1.00 & -0.16 & 0.20 & 0.04 & 0.04 & 0.03 \\
    \hline
    0.35 & $0.815 \pm 0.020$ & 0.59 & -0.16 & 1.00 & -0.24 & 0.21 & -0.14 & 0.12 \\
    \hline
    0.55 & $0.664 \pm 0.024$ & 0.10 & 0.20 & -0.24 & 1.00 & -0.50 & 0.32 & -0.11 \\
    \hline
    0.90 & $0.790 \pm 0.066$ & 0.15 & 0.04 & 0.21 & -0.50 & 1.00 & -0.66 & 0.25 \\
    \hline
    1.50 & $0.18 \pm 0.11$ & -0.05 & 0.04 & -0.14 & 0.32 & -0.66 & 1.00 & -0.34 \\
    \hline
    2.00 & $0.46 \pm 0.22$ & 0.07 & 0.03 & 0.12 & -0.11 & 0.25 & -0.34 & 1.00 \\
    \hline
    \end{tabular}
    ~
\begin{tabular}{|c|c|cccc|}
    \hline
    \multicolumn{6}{|c|}{CMB compressed likelihood}\\
    \hline
    Variable & Mean $\pm$ error & \multicolumn{4}{c|}{Correlation matrix}\\
    \hline
    $R$ & $1.7508 \pm 0.0041$ & 1.00 & 0.35 & -0.64 & -0.73 \\
    \hline
    $l_a$ & $301.544 \pm 0.084$ & 0.35 & 1.00 & -0.24 & -0.28 \\
    \hline
    $\omega_b$ & $0.02235 \pm 0.00015$ & -0.64 & -0.24 & 1.00 & 0.45 \\
    \hline
    $n_s$ & $0.9644 \pm 0.0043$ & -0.73 & -0.28 & 0.45 & 1.00 \\
    \hline
\end{tabular}
    \caption{\textit{Left.} Pantheon+ SNe data compressed in data driven measurements of $1/E(z)$. \textit{Right.} CMB compressed variables. We show in \cref{fig:lcdm_compressed} and \cref{fig:w0wa_SNe_compressed} that a Gaussian likelihood with the mean and covariance provided here is good enough to recover the posterior distribution, in both cases.}
    \label{tab:SNe_CMB_compressed}
\end{table*}
\end{center}

\section{Analytic approximation for initial conditions}\label{sec.symbolic}

The history of a thawing quintessence model is uniquely determined by the parameters of the potential, ($V_0$, $m^2$) and the initial value of the scalar field $\varphi_{\rm ini}$ (recall that we are assuming that ${\dot \varphi}_{\rm ini}=0$) at some early time. This will determine the final fractional energy density in the scalar field $\Omega_{\varphi}=1-\Omega_{\rm m}$. We are interested, however, in choosing $\Omega_\varphi$, or more specifically, $\Omega_{\rm m}$. This means that a set of ($V_0$, $m^2$) and $\Omega_{\rm m}$ should uniquely determine $\varphi_{\rm ini}$. In practice, this is done  by  using a root-finding algorithm in \texttt{hi\_class}, which finds the correct $\varphi_{\rm ini}$. This is a time consuming process and sufficiently onerous that it makes an MCMC inference prohibitively slow. It is preferable to have an accurate predictor of $\varphi_{\rm ini}$ which can greatly reduce evaluations of \texttt{hi\_class}. 

For both computational efficiency and for interpretability, we wish to find symbolic approximations for $\varphi_{\rm ini}$ as a function of $V$, $m^2$, and the background cosmological parameters $\Omega_{\rm m}$ and $h$. 
To obtain such an expression, we utilize the supervized machine learning technique of Symbolic Regression (SR) \citep{Kronberger_2024}: an automated search through function space to find simple mathematical expressions which well approximate a given dataset.

We use the genetic programming based SR code \textsc{operon} \citep{Burlacu_2020}, which uses natural selection inspired processes such as crossover (breeding) and mutation to evolve an initial population of functions over time, where well-fitting and simple expressions are preferentially kept and worse-fitting or overly-complex functions are discarded.
As such, after several iterations, a population of expressions which well approximate the dataset emerge. We choose \textsc{operon} over other SR codes due to its strong performance in benchmark tests \citep{LaCava_2021,Burlacu_2023} and as it has already been demonstrated to be effective in cosmological and astrophysical studies \citep{Bartlett_2023_Pofk,Bartlett_2024_syren,Russeil_2024,AbdusSalam_2024}.

To obtain a symbolic approximation, we first need a training set. It is easier to obtain $m^2$ as a function of $\varphi_{\rm ini}$ and the other parameters than vice versa, hence we generated $\sim$ 10,000 points on a Latin hypercube in the range 
$\varphi_{\rm ini} \in [0.1, 10]$, $V_0 \in [-10,10]$, $\Omega_{\rm m}\in [0.01, .99]$ and $h \in [0.2,1.0]$.
We then obtain the corresponding value of $m^2$ for each set of parameters by using \texttt{hi\_class}. Any points which do not provide converged solutions are removed.
We use 90\% of these data for training and the remainder as a validation set.

Due to the large dynamic range of $\varphi_{\rm ini}$, we choose to fit for $\log \varphi_{\rm ini}$ instead of $\varphi_{\rm ini}$ directly. We apply a multi-objective optimization scheme, where we attempt to minimize both the mean absolute error of $\log \varphi_{\rm ini}$ (and thus the fractional error on $\varphi_{\rm ini}$) and the model length, which is approximately equal to the number of symbols appearing in the expression. 
\textsc{operon} relies on the concept of $\epsilon$-dominance \citep{Laumanns_2002}, such that these objective values are considered equal in the search if they fall within some hyper-parameter $\epsilon$ of each other. After some experimentation, we find that $\epsilon = 10^{-2}$ is an appropriate choice for balancing accuracy and simplicity.

We fit $\log \varphi_{\rm ini}$ as a function of $V$, $m^2$, $\Omega_{\rm m}$ and $h$, where we allow symbolic expressions containing the operators $\splitatcommas{+, -, \times, /, \sqrt{\cdot}, {\rm pow}, \log, \exp}$ and free parameters, which are optimized \citep{Kommenda_2020} using the Levenberg–Marquardt algorithm \citep{Levenberg_1944,Marquardt_1963}.
We run \textsc{operon} for up to 24 hours, allowing a maximum model length of 140. 
We visually inspect the candidate solutions to obtain one which we deem to balance accuracy and simplicity, enforcing that our solution has a mean absolute error of at least 1\% on both the training and validation sets.
These models are often over-parameterized, so we manually remove superfluous parameters and attempt to round those which are close to integers, zero or simple rational numbers to try to simplify the expression.

Given the very different behaviors of the $m^2<0$ and $m^2>0$ branches, we choose to use a different expression for each.
After some manipulation, for $m^2<0$, we obtain the following expression of length 23 for the initial $\varphi$:
\begin{equation}
    \log \varphi_{\rm ini} \approx m^2  e^{b_0 h} 
    + \frac{1}{2} \log \left( \frac{b_1 h^2 \left( 1 - \Omega_{\rm m} \right) - V_0}{m^2} \right),
\end{equation}
where $b_0\approx-1.5$ and $b_1 \approx 3.0$. This approximation gives an error on the training and test sets of $8\times10^{-3}$ and $9\times10^{-3}$, respectively, so is sufficiently accurate for our purposes.

We find that the functional form for $m^2>0$ has to be more complex, with an expression of length 48 fulfilling our selection criteria. This is given by
\begin{widetext}
    \begin{equation}
    \begin{split}
        \log & \varphi_{\rm ini} \approx
        c_{0} 
        + c_{1} \left( \frac{\Omega_{\rm m}}{c_{2} V_0 + c_{3} h + e^{c_{4} \Omega_{\rm m}}} + c_{5} V_0 \right)^{-1} \\
        &+ c_{6} \left(\frac{\Omega_{\rm m}}{c_{7} V_0 + c_{8} h + e^{c_{9} \Omega_{\rm m}}} - c_{10} m^{2} + c_{11} \left( c_{12} \Omega_{\rm m}  - m^{2}\right)^{c_{13}} \left(\Omega_{\rm m} + c_{14} V_0 - c_{15} m^{2}\right)\right) 
        \left( \left( c_{16} \Omega_{\rm m}\right)^{c_{17} h^{c_{18}}} - e^{c_{19} V_0} \right)^{-1},
    \end{split}
\end{equation}
\end{widetext}
where
$\bm{c} = \{ \splitatcommas{0.006, 0.0558, 0.8222, -3.1121, 1.4964, -1.2683, -1.775, 0.9655, -3.3049, 1.3418, -0.0816, -1.1722, 0.0291, -0.3267, 0.2304, -0.9416, 54.7118, 0.3744, 0.6663, -0.5982} \}$.
We find that this expression gives a mean absolute error of 0.01 on the training set, and $8\times10^{-3}$ on the test set, so is able to give a percent-level approximation for $\varphi_{\rm ini}$, so is again sufficiently accurate for our purposes.
The small errors in our approximations are corrected for by the root-finding algorithm in \texttt{hi\_class}, but this requires fewer iterations than if we used a random starting point for the optimizer.

\section{$m^2$ prior}\label{sec.m2}
\begin{figure}
    \centering
    \includegraphics[width=\columnwidth]{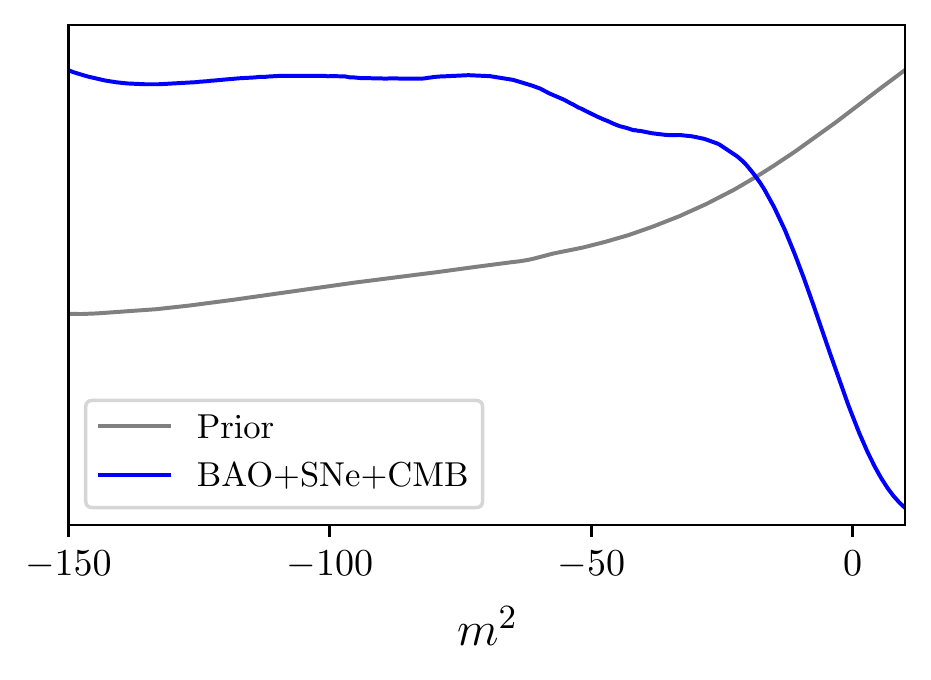}
    \caption{1D-posterior distribution for $m^2$ from the combination of DESI BAO data \cite{DESI:2024mwx}, Pantheon+ SNe data \cite{Scolnic:2021amr}, and CMB data \cite{Planck:2018vyg, Planck:2019nip, ACT:2023kun, ACT:2023dou} (blue) and the effective prior imposed (gray).}
    \label{fig: m2 prior}
\end{figure}

As discussed in \cref{Sec: Dynamical Dark Energy}, an important aspect of our work is ensuring that we have effective flat priors in the parameters that we sample in the inference. This is particularly relevant for $m^2$ if we want to compare in a meaningful way thawing quintessence with $\Lambda$CDM. We show in \cref{fig: m2 prior} the effective prior and the 1D-posterior distribution from the combination of all data (DESI BAO, Pantheon+ SNe Ia and CMB).  As mentioned in \cref{Sec: Dynamical Dark Energy}, there is a degeneracy between $m^2$ and $V_0$ that enhances the prior on the $m^2>0$ branch slightly. However, the prior on $m^2$ is effectively flat across the range of masses considered. We find that this extra volume at positive $m^2$ does not drive the constraints and data actually disfavors this regime.

\section{Full posterior distribution for thawing quintessence}\label{sec.fullposterior}
\begin{figure*}[htbp]
    \centering
    \begin{minipage}[b]{0.9\textwidth}
        \centering
        \includegraphics[width=\textwidth]{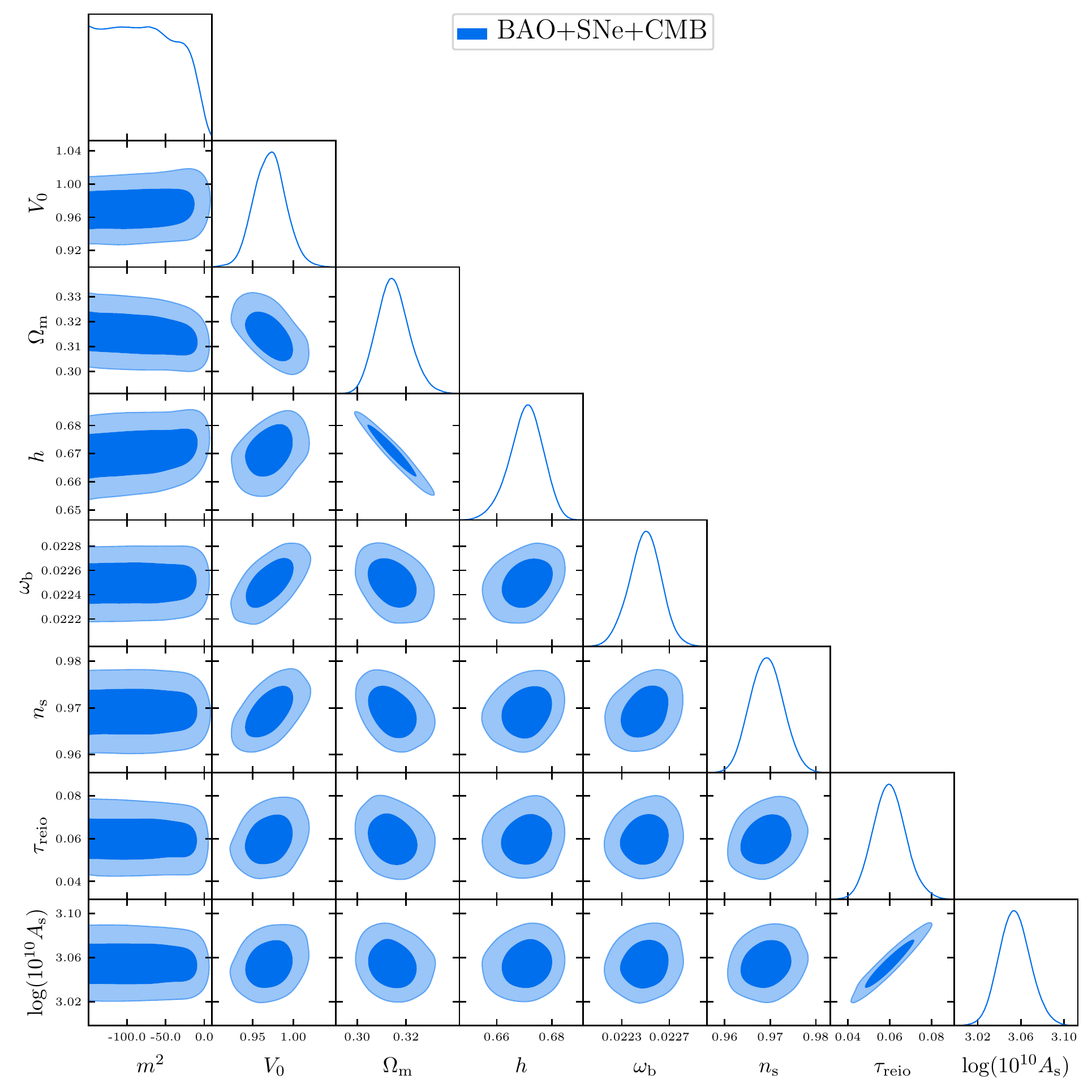}
        \label{fig:subfig1}
    \end{minipage}
    \caption{68\% and 95\% C.L. marginalized posterior distributions for the full cosmological parameter space of thawing quintessence from the combination of DESI BAO data \cite{DESI:2024mwx}, Pantheon+ SNe data \cite{Scolnic:2021amr}, and CMB data \cite{Planck:2018vyg, Planck:2019nip, ACT:2023kun, ACT:2023dou}.
    }\label{Fig:mcmc_m2_full}
\end{figure*}

\cref{Fig:mcmc_m2_full} gives the full corner plot featuring all parameters varied as per Table.~\ref{priors} when constraining thawing quintessence in terms of the parameters $(V_0, m^2)$.

\clearpage
\bibliography{refs}

\end{document}